\documentclass[sigconf]{acmart}

\AtBeginDocument{%
  }

\copyrightyear{2025}
\acmYear{2025}
\setcopyright{cc}
\setcctype{by}
\acmConference[FAccT '25]{The 2025 ACM Conference on Fairness, Accountability, and Transparency}{June 23--26, 2025}{Athens, Greece}
\acmBooktitle{The 2025 ACM Conference on Fairness, Accountability, and Transparency (FAccT '25), June 23--26, 2025, Athens, Greece}\acmDOI{10.1145/3715275.3732046}
\acmISBN{979-8-4007-1482-5/2025/06}

\usepackage{tabularx}
\usepackage{booktabs}
\usepackage{multirow}

\begin{document}

\title[Measuring machine learning harms from stereotypes]{Measuring Machine Learning Harms from Stereotypes Requires Understanding Who Is Harmed by Which Errors in What Ways}

\author{Angelina Wang}
\affiliation{%
  \institution{Stanford University, Cornell Tech}
    \country{USA}
  }
\email{angelina.wang@cornell.edu}

\author{Xuechunzi Bai}
\affiliation{%
  \institution{University of Chicago}
    \country{USA}
  }

  \author{Solon Barocas}
\affiliation{%
  \institution{Microsoft Research}
    \country{USA}
  }

  \author{Su Lin Blodgett}
\affiliation{%
  \institution{Microsoft Research}
    \country{USA}
  }
\begin{abstract}
Despite a proliferation of research on the ways that machine learning models can propagate harmful stereotypes, very little of this work is grounded in the psychological experiences 
of people exposed to such stereotypes. We use a case study of gender stereotypes in image search to examine how people react to machine learning errors. First, we use surveys to show that not all machine learning errors reflect stereotypes nor are equally harmful. Then, in experimental studies we randomly expose participants to stereotype-reinforcing, -violating, and -neutral machine learning errors. We find stereotype-reinforcing errors induce more experiential harm, while having minimal impact on participants' cognitive beliefs, attitudes, or behaviors. This experiential harm impacts participants who are women more than those who are men. However, certain stereotype-violating errors are more experientially harmful for men, potentially due to perceived threats to masculinity. We conclude by proposing a more nuanced perspective on the harms of machine learning errors---one that depends on who is experiencing what harm and why.
\end{abstract}

\ccsdesc[500]{Social and professional topics~Gender}
\ccsdesc[500]{Computing methodologies~Object recognition}

\keywords{stereotypes, machine learning fairness, harms from biases}


\maketitle

\section{Introduction}

Over the past decade, researchers have demonstrated that machine learning models run the risk of learning stereotypical associations. For example, natural language processing models have been shown to associate women with homemakers and men with programmers \citep{bolukbasi2016embeddings}, while computer vision models have been shown to associate women with shopping and men with driving \citep{zhao2017men}. These associations can cause machine learning models to make systematic errors, mistakenly reporting, for instance, that female doctors are nurses and that male nurses are doctors \citep{selvaraju2017grad}. A rich literature has developed offering many more such examples, spurring calls to address the risk that machine learning models might make mistakes that perpetuate harmful stereotypes.

\begin{figure*}[t!]
    \centering
    \includegraphics[width=0.9\textwidth]{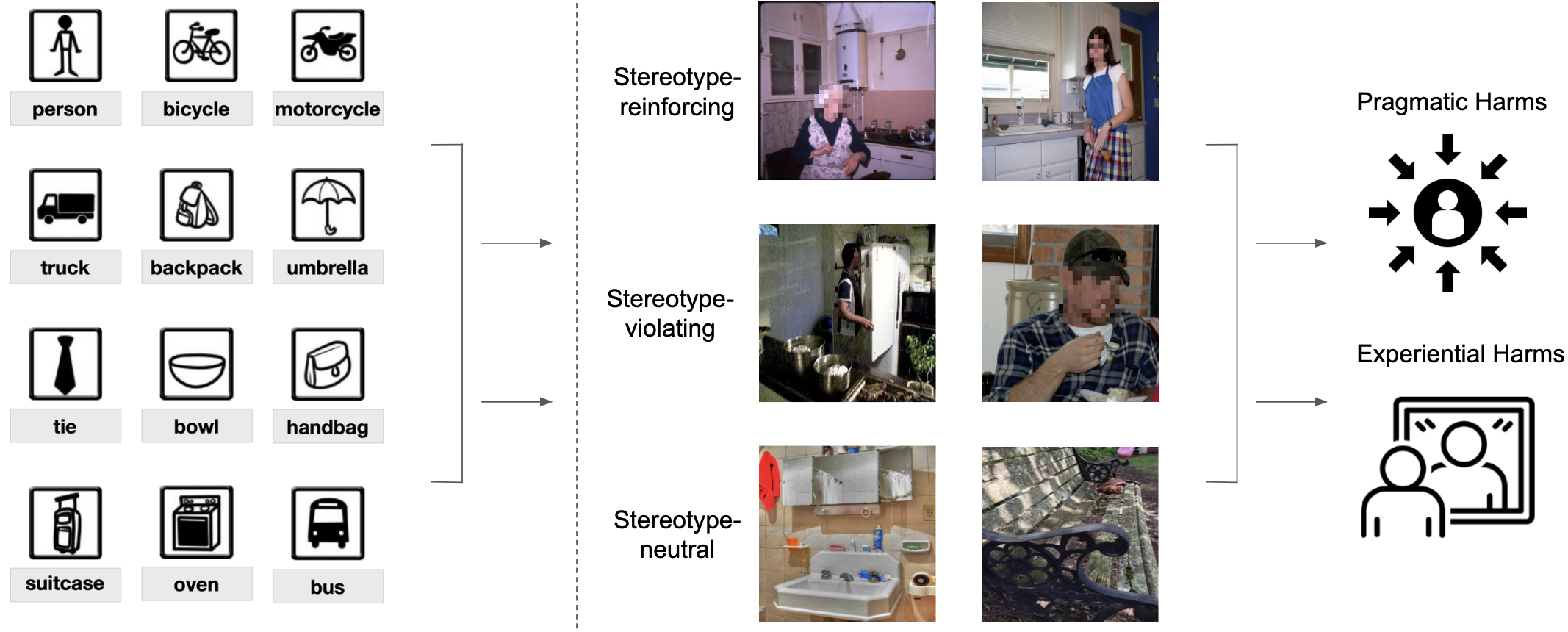}
    \caption{
    \textbf{Summary of our studies.} The left side represents studies 1 and 2, where we ask human participants to mark which of the relevant objects in our application are stereotypically associated with which gender groups, as well as to qualitatively explain why that is and why it is harmful or not. The right side represents studies 3 and 4 where we randomly expose participants to machine learning errors which are stereotype-reinforcing, stereotype-violating, or stereotype-neutral---as determined by the annotations from study 1. Then, we measure two forms of harm we introduce: pragmatic (measurable changes in someone's cognitive beliefs, attitudes, or behaviors toward the group being stereotyped) and experiential (self-reports of negative affect). The images shown are examples of misclassifications of \texttt{oven} where stereotype-reinforcing errors are when it is falsely predicted on a woman, stereotype-violating when on a man, and stereotype-neutral when no gender stereotypes are invoked. These images are shown to participants on a search result page of \texttt{oven}.}
    \label{fig:error_types}
\end{figure*}

Unfortunately, portions of this literature have suffered from three main limitations that impede effective intervention. First, while some research has asked crowd workers to annotate when errors invoke stereotypes or has drawn on pre-existing inventories of stereotypes~\citep{cao2022theory,bolukbasi2016embeddings, sotnikova2021stereotype, caliskan2017semantics, bianchi2023generation, wan2023kellywarm}, machine learning researchers often rely on their own moral intuitions to determine which categories of associations to investigate (e.g., associations between gender and occupation) and to judge which specific associations (e.g., the association between women and nursing) are stereotypes. As a result, certain categories of associations that broader populations might perceive as stereotypical have not been investigated and many discovered associations within these categories are assumed to be stereotypical, even if broader populations would not perceive them to be so. For example, researchers identified an object recognition model as reproducing stereotypes because it amplifies the degree to which labels for kitchen items like ``knife,'' ``fork,'' and ``spoon'' are incorrectly assigned to photos featuring women, and labels for technology-related items like ``keyboard'' and ``mouse'' are incorrectly assigned to photos featuring men \citep{zhao2017men}. However, these intuitions might not always be shared by the broader population; indeed, as we'll show later in this paper, some of these associations are not seen as invoking stereotypes among a sample of people in the United States.

Second, prior work has tended to ignore whether errors reinforce or violate stereotypes, treating each type of error as equally harmful. Researchers often resort to treating all spurious correlations between objects and specific social groups as harmful, even if correlations violate rather than reinforce stereotypes. 
As a result, it remains unclear the degree to which the harmfulness of errors depends on whether and how these errors invoke stereotypes. And in fact, as we'll show, stereotype-reinforcing errors and stereotype-violating errors have disparate, sometimes contrasting, impacts on those harmed.


Third, while stereotypes are frequently invoked to explain why some machine learning errors are more harmful than others~\citep{barlas2021see, bhaskaran2019secretaries, abbasi2019stereotyping, revisetool_extended}, little has been done to establish the actual impact that these stereotype-aligned errors have on people in practice. Rather than measuring the harms that stereotypes bring about and identifying the particular mechanisms by which they do so, much of the literature simply presumes that machine learning stereotypes have negative impacts on the people so stereotyped or on society more generally. While some prior work has found that exposure to gender-biased image search results can lead both to more biased estimations of the representation of different gender groups in certain occupations and to a decreased sense of belonging ~\citep{kay2015search,metaxa2021image}, there remains a paucity of empirical work that seeks to measure the psychological and practical effects of exposure to such errors and to characterize the nature of these harms. In fact, not all stereotypes are harmful~\cite{beeghly2015stereotype, beeghly2025stereotype}---we hold stereotypes that children are bad drivers, yet generally do not see this as a harmful stereotype.
Recent FAccT work has pointed out how measures of representational harm like stereotyping have tended to be overly focused on behaviorist definitions, neglecting those which affect cognitive and affective states~\cite{chien2024behondbehaviorist}. In our work, we study these overlooked dimensions based on the psychological notion that stereotypes themselves are cognitive beliefs which may manifest in different ways.


In this paper, we draw on the psychological literature on social stereotypes to overcome each of these limitations, performing a series of empirical studies in which human participants are exposed to machine learning errors (Fig.~\ref{fig:error_types}). As a concrete application to ground these studies, we focus on gender stereotypes in the machine learning task of object recognition, which is now a common feature in photo management software like Apple Photos and Google Photos.

\begin{figure*}[t!]
    \centering
    \includegraphics[width=0.8\textwidth]{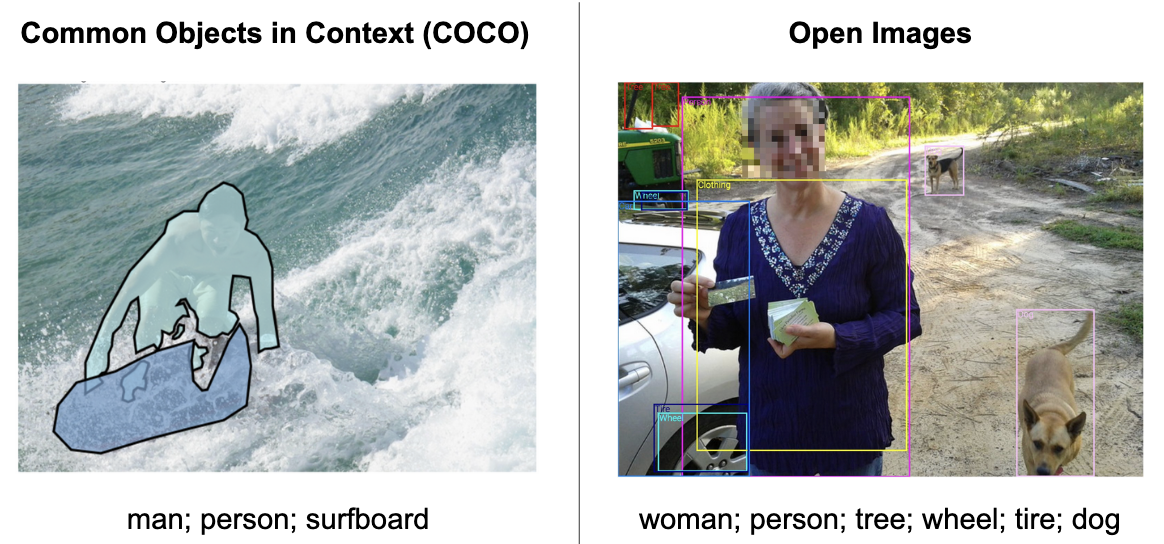}
    \caption{\textbf{COCO and Open Images object recognition datasets.} We use two commonly used image recognition datasets to represent the application of a photo search engine. Both datasets contain annotations for perceived binary gender expression of the people in the images and the objects present in each image. The left panel shows one example figure from COCO annotated with \texttt{surfboard}. The right panel shows one example figure from Open Images annotated with objects such as \texttt{tree} and \texttt{dog}. 
    }
    \label{fig:datasets}
\end{figure*}

To overcome the first limitation of which associations people view as stereotypical, we draw on the psychological literature which suggests that the space of stereotypical associations is potentially much broader than what has been examined in the machine learning literature~\citep{nicolas2022spontaneous, hilton1996stereotypes, fiske2002scm, ellemers2018gender}. Hence, we make no a priori decisions about the categories of association worth investigating. We expand our scope of analysis beyond commonly focused-on categories---for example, occupations or activities---to include all objects that might appear in an image. We do so via two studies. In the first (Study 1), we present participants with images from popular computer vision datasets that include a range of objects (COCO~\citep{lin2014coco} and OpenImages~\citep{kuznetsova2020openimages} datasets (Fig.~\ref{fig:datasets})), and ask whether these objects are stereotypically associated with different gender groups. To better understand why certain objects are seen as possessing stereotypical associations, we conduct a qualitative analysis (Study 2) on participants' open-ended responses to the questions in Study 1.\looseness=-1


Then, to overcome the second and third limitations of imprecision about which kind of stereotypical errors may cause what kind of harm, we causally assess the effect that exposure to these stereotypes have on people in practice. We focus on two kinds of potential harm: pragmatic and experiential. 
Research in psychology describes stereotypes as cognitive beliefs in people's minds, which can have an influence on attitudes (i.e., prejudice) and behaviors (i.e., discrimination)~\citep{lippmann1922public,katz1933racial, allport1954nature, hilton1996stereotypes, hamilton2014stereotypes}.
Beyond these beliefs, attitudes, and behaviors, members of the stereotyped social group may also feel disrespected, demeaned, or otherwise discounted. Such experiences can be thought of as dignitary harms that bring about negative affect in members of the group so stereotyped~\citep{katzman2023taxonomy}. While dignitary harms are frequently treated as less important than the concrete effects of changes in beliefs, attitudes, and behaviors, they may impose a substantial emotional toll, akin to 
microaggressions~\citep{rini2020microaggression}.
We therefore introduce a distinction between \textit{pragmatic harms}, which involve measurable changes in someone’s cognitive beliefs, attitudes, or behaviors toward the group being stereotyped, and \textit{experiential harms}, which involve self-reports of negative affect.

Additionally, we differentiate between errors that are stereotype-reinforcing, stereotype-violating, or stereotype-neutral. This is because errors that invoke stereotypes may do so in different ways and may therefore have different effects. For example, an error associating women with homemakers reinforces a gender stereotype and may make women feel their agency is devalued. In contrast, an error associating men with homemakers violates a gender stereotype and may have little effect on men's self-perceived agency and instead threaten their masculinity~\citep{shnabel2016help}.

With the above considerations in place, we examine the extent to which participants are harmed by different types of stereotypical errors in an image recognition task. Concretely, we randomly assign participants to synthesized image search result pages which contain different kinds of errors, then measure pragmatic harms through monitoring changes in cognitive, attitude, and behavior measures, and experiential harms by collecting negative affective states reported by participants.

We find little immediate causal effect of pragmatic harms, but sizable evidence that stereotype-reinforcing errors are experientially harmful---a finding that is more pronounced among participants who identify as women compared to those who identify as men (Study 3). We find that while the stereotyped group (e.g., women) generally finds it more experientially harmful for the error to reinforce rather than violate stereotypes, this is not true when it comes to clothing-related items typically associated with women (e.g., \texttt{cosmetics}, \texttt{necklaces}) being misclassified on men. Here, we see a backlash towards violations of the norms around gender presentation: Men tend to find these misclassifications of, e.g., \texttt{cosmetics}, more harmful on men rather than women. This last observation calls into question the idea that it is always normatively desirable to reduce errors perceived as more harmful due to their relationship to stereotypes (Study 4).

By bringing greater clarity to different types of machine learning errors based on their relationship to a stereotype and embracing the rich psychological experiences of such errors, we urge researchers and practitioners to more carefully consider different kinds of machine learning errors, potential harms, and the relevant relationships between them. Investigating the psychological experiences that people have when encountering machine learning errors is critical to understanding the potential harm of a system, and in turn, mitigating it. 

The paper is organized as follows: We first discuss related work, then give context for our case study on gender stereotypes in photo search. Next, we report four human studies, and conclude with a discussion of our results.

\section{Related Work}
Stereotyping is a common (often implicit) operationalization of ``bias'' that is studied in the literature on fairness in machine learning~\cite{barlas2021see, bhaskaran2019secretaries}.
\citet{abbasi2019stereotyping} mathematically formalize stereotyping and show how models trained on stereotyped data can lead to unfair outcomes.
Bias amplification is a different statistical notion that rests on the idea that any amplification of an existing bias is undesirable~\cite{zhao2017men, wang2021biasamp, wang2019balanced, hall2022biasamp}. We show in our work that these approximations of the type of error that is more harmful is often poorly correlated with human judgment, and provide a detailed comparison in Appendix~\ref{app:biasamp}.
Work in natural language processing considers stereotypes reflected in word embeddings~\cite{caliskan2017semantics, bolukbasi2016embeddings} and across many tasks, such as language modeling~\cite{nangia-etal-2020-crows,cao2022theory,jha-etal-2023-seegull}, machine translation~\cite{savoldi-etal-2021-gender}, and question answering~\cite{parrish-etal-2022-bbq}, with critiques pointing out that benchmarks aimed at measuring stereotyping may fail to meaningfully conceptualize and operationalize stereotyping~\cite{blodgett2021salmon}.
Though researchers have proposed different operationalizations of stereotypes~\cite{dev2022biasnlp, gallegos2023survey, czarnowska2021socialbiasnlp}, there is generally a large reliance on occupation data from the American Bureau of Labor Statistics---e.g., WinoBias~\cite{zhao2018winobias}---as a data source for stereotypes. One large limitation of this usage (in addition to it only representing American occupation data) is that it only captures descriptive stereotypes, e.g., overrepresentations of groups in an occupation, and misses prescriptive stereotypes, e.g., beliefs about what occupations people of different groups should be in; these two types can often differ in practice~\cite{lopezsaez2014descriptiveprescriptive, burgess1999descriptiveprescriptive}. More recent work has addressed some of these issues by proposing a community-driven approach to collecting stereotypes that are uniquely present in India~\cite{dev2023stereotypeindia, jha2023seegull}.

\textbf{Psychology of Stereotypes and their Consequences.} Psychology research describes stereotypes as cognitive beliefs, which can have an influence on attitudes (i.e., prejudice) and behaviors (i.e., discrimination)~\citep{lippmann1922public,katz1933racial, allport1954nature, hilton1996stereotypes, hamilton2014stereotypes}. For example, people may have \textit{cognitive beliefs} that women are more warm but less competent than men, and thus \textit{express} protective attitudes and pity for women~\citep{glick1996asi, cuddy2007bias}. People then \textit{behave} in ways that maintain women's warmth and discount their competence, such as being less likely to promote women to leadership positions~\citep{fiske2002scm,ellemers2018gender}. 
Therefore, stereotypes of certain social groups can prompt shifts in attitudes and behaviors that can ultimately harm the stereotyped group. We draw on these distinctions to formulate our measure of pragmatic harm.\looseness=-1

\textbf{Image Search. }
We concretize our task of object recognition by constructing human studies involving the application of image search. There is a large body of work on biases in image search engines~\cite{otterbacher2017competent, otterbacher2018sexism, kay2015search, metaxa2021image, vlasceanu2022internet, guilbeault2024amplify}, as well as search algorithms and media systems more broadly~\cite{noble2018oppression, graves1999television, introna2000web, diaz2008goggles}. 

Most findings on image search engines have focused on auditing the search engine itself for biased results~\cite{otterbacher2017competent, otterbacher2018sexism, kay2015search, metaxa2021image, vlasceanu2022internet, guilbeault2024amplify}, leaving more speculative the implications of these results for feeding back into the stereotypes people hold. However, some components of this work also attempt to understand the influence that biased search results can have on users. 
\citet{kay2015search} and \citet{metaxa2021image} both find that exposure to gender biased image search results can influence an individual's opinion on the gender representation of that occupation (accounting for 7\% of a person's opinion in the former and roughly 5\% increases, with great variance, when representation shifts from 10\% to 50\% to 90\% in the latter).
\citet{metaxa2021image} also find an effect of .146 (95\% CI [-.16, .45]) on a 7-point Likert scale for the impact of gender representation on feelings of an occupation's inclusivity, with no effect for interest in an occupation. Notably, these feelings of inclusivity depend on the gender of the participant. \citet{vlasceanu2022internet} take a different approach by studying occupations (e.g., peruker, lapidary) for which there are very few preconceived stereotypes.  
Our work focuses on the reinforcement of existing stereotypes, rather than trying to induce new ones.
Recent work by \citet{guilbeault2024amplify} conducts a large-scale study of social categories including occupations and roles, and measures psychological effects such as through the Implicit Association Test.
Unlike much of this existing work, we do not audit an existing image search engine, instead focusing on downstream concerns that may arise from biases in the search results. We also expand the scope of what is a stereotype (e.g., objects).

\textbf{Crowdsourcing Opinions on Harm. }
We solicit our labels of stereotypes and harm from crowdworkers, similar to how
prior work has elicited stereotypes~\cite{bolukbasi2016embeddings, sotnikova2021stereotype} and fairness notions in different contexts~\cite{yaghini2021human, srivastava2019perception, grgichlaca2018perception}. \citet{woodruff2018qualitative} specifically focus on the opinions of individuals from marginalized communities who are likely to be adversely affected by algorithms. Unlike this existing work, however, we further attempt to measure the psychological effect of exposure to machine learning errors, rather than relying exclusively on self-reported opinions. In particular, we measure changes in participants' beliefs, attitudes, and behaviors, as well as the extent to which they feel that they have been \emph{personally} harmed by these errors.

\section{Case Study: Gender Stereotypes in Photo Search}
We explore a popular task in machine learning known as object recognition (i.e., classifying the objects present in an image). To make it concrete for our human studies, we use it in the context of a smart phone's photo search engine, and examine gender stereotypes. Specifically, we consider one type of machine learning error called a false positive: when an object is predicted to be present in an image when it is in fact not there. This causes the image with a false positive to be wrongly surfaced on an image search results page.\footnote{We note that false negatives are subsumed in this setting because enough false positives will crowd out the results page and ultimately have a similar effect as false negatives on images of the gender that does not have false positives.} We study the \emph{effect} of the misclassification, not why the model may have made the mistake, or what the participant thinks is the reason the model made the mistake. 
We use a mixture of qualitative and regression analyses to report our findings. For our within-subjects surveys, we regress with a mixed-effect model whose parameter estimations are adjusted by the group random effects for each individual. We report the coefficients from our regression analyses, which represent the effect size of that independent variable. Further details on our methods and analysis are in Appendix~\ref{app:methods}, and code is at \url{https://tinyurl.com/mlstereotypes-code}.
To protect participants' privacy, data is available upon request.
All studies are approved by our institution IRB.
Studies 1 (\url{https://tinyurl.com/mlstereotypes-study1}), 3 (\url{https://tinyurl.com/mlstereotypes-study3pt1}, \url{https://tinyurl.com/mlstereotypes-study3pt2}), and part of Study 4 (\url{https://tinyurl.com/mlstereotypes-study4})
are pre-registered on OSF, while Study 2 is exploratory. 

\section{Study 1: Distinguishing which machine learning errors reflect social stereotypes}

\subsection{Method}

To understand the social stereotypes held by American society relevant to our machine learning task,
we first elicit human judgments (\textit{N} = 80) on Common Objects in Context (COCO)~\citep{lin2014coco}. COCO has 80 objects and perceived binary gender expression of pictured people annotated across the images~\citep{zhao2021caption}. In the study, we ask the participants whether each object (e.g., \texttt{keyboard}, \texttt{zebra}) is stereotypically associated with men, women, or neither. We recruited U.S.-based participants on Cloud Research with filters for those who had at least 50 HITs approved, and a HIT approval rate of 98\%. We paid a rate of \$15/hour and explicitly recruited equal numbers of participants who identified as men and as women.

\subsection{Results}
\begin{figure*}[t!]
    \centering
    \includegraphics[width=0.99\textwidth]{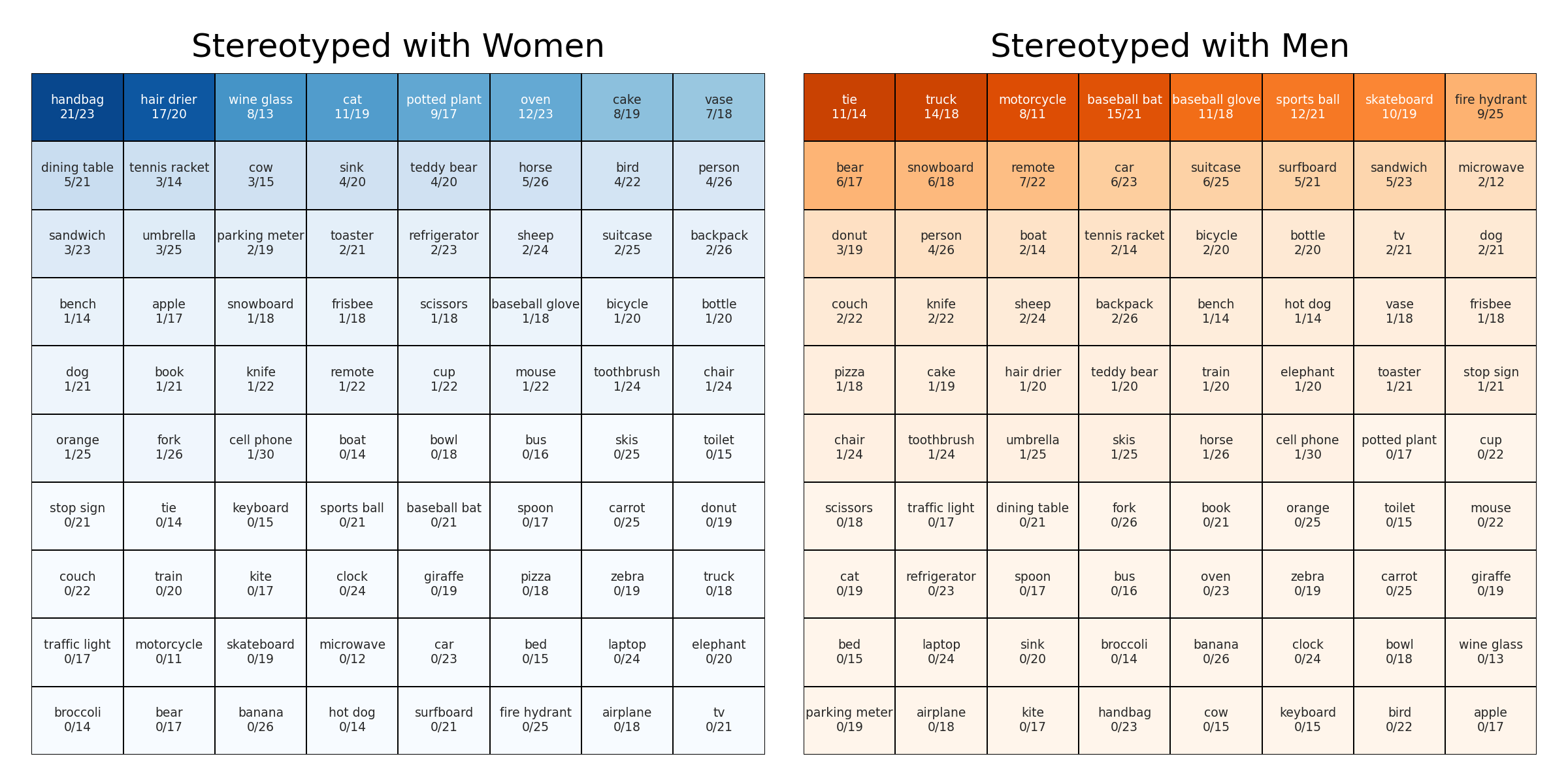}
    \caption{\textbf{Study 1 Object Results.} Detailed participant responses for each of the 80 objects in COCO dataset. Fraction indicates number of participants asked about each object who marked it as stereotypically related to the gender group of women or men.}
    \label{fig:heatmap}
\end{figure*}

As expected, not all objects reflect gender stereotypes. This is already in contrast to a somewhat common assumption in ML fairness research that \emph{any} difference between groups is an amplification of a stereotype~\citep{blodgett2021salmon}. 
Among 80 objects, 13 objects are marked as stereotypes by more than half of the participants (Figs.~\ref{fig:heatmap}). Some examples of stereotypically gendered objects are \texttt{handbag} with women, \texttt{wine glass} with women, \texttt{tie} with men, and \texttt{truck} with men. Among the remaining objects, 18 objects (e.g., \texttt{keyboard}, \texttt{carrot}, \texttt{traffic light}) are marked by zero participants as stereotypes with any gender group, challenging prior assumptions on what is seen as a stereotype~\citep{zhao2017men}.
If an object was marked to be a stereotype, we also asked participants whether they believed it was harmful in the abstract. Complete results are in Appendix~\ref{app:study1_results}, but we use these initial findings to select experimental stimuli in subsequent studies.
In Study 3a the stereotype-reinforcing condition includes women and \texttt{oven} (marked to be most harmful), women and \texttt{hair dryer} (marked to be least harmful), and the associated control conditions include women and \texttt{bowl}, women and \texttt{toothbrush}. In Study 3b we also include in the stereotype-reinforcing conditions of men and \texttt{baseball glove} (marked to be more harmful) and men and \texttt{necktie} (marked to be less harmful) with the control conditions of men and \texttt{bench} and men and \texttt{cup}.

\section{Study 2: Plurality of stereotypes and harms with image recognition objects}

\subsection{Method}
Next, we report qualitative analyses on open-ended responses from participants' annotations, where they explain why certain objects are seen as stereotypes and harmful or not. 
While prior work in gender stereotypes has often focused on social roles and traits~\citep{glick1996asi, eagly1987socialrole}, our data provides insights as to how objects (e.g., \texttt{oven}, \texttt{hair dryer}) can also be associated with stereotypes. This is an important departure because it expands the scope of machine learning tasks for which stereotypes are relevant beyond its current more narrow framing. Specifically, when a participant from Study 1 responds that an object is a stereotype, we follow up and ask: ``Please describe in 1-2 sentences a) why you marked the above as a stereotype, and b) why you found it to be harmful or not.''

\subsection{Results}
One of the authors coded the responses for why an object is a stereotype into six categories. The coding was done inductively, then thematically clustered to form the six categories. The most prevalent reasons were: descriptive (45\%), e.g., for \texttt{handbag} and women: ``women are often seen wearing handbags and buying them''; occupation/role (22\%), e.g., for \texttt{oven} and women: ``women are stereotyped to always be in the kitchen cooking while the men go out and work''; trait (11\%), e.g., for \texttt{chair} and men: ``sometimes men would be seen as coming home and just being lazy and lounging in their chair.'' The full analysis is in Appendix~\ref{app:study2_results}. It is interesting to note that an object's association to a stereotype is frequently mediated by its connection to a role or trait, which are the more common sites of inquiry when it comes to stereotypes~\citep{eagly1987socialrole, fiske2002scm, ellemers2018gender}. We also found that associations between a group and an object can exist through a number of paths. For example, explanations for stereotypical associations between cats and women include: ``cat lady,'' ``women are called \textit{kitten},'' ``women like cats more than dogs,'' ``cats are a feminine animal,'' and ``women are called \textit{cougars}.''

When asked why a stereotype was harmful or not, many respondents simply reiterated that the object was a stereotype. Dropping those responses, one of the authors coded the free responses of why a stereotype was marked to be harmful into seven categories, with the top three being: proscriptive (40\%), e.g., for \texttt{dining table} and women: ``it makes it looked down upon if a man cooks dinner''; prescriptive (26\%), e.g., for \texttt{dining table} and women: ``I think it puts women in a box that says they must prepare dinner''; negative trait (13\%), e.g., for \texttt{handbag} and women: ``it is harmful because it implies that women cares more about looks and their appearance.'' The remaining response categories are in Appendix~\ref{app:study2_results}.
There seems to be a disparity in responses based on the participant's gender regarding whom they believe is harmed. When women specify which of the men group or women group are harmed, they say it is the women group 79\% (95\% CI [.67, 88]) of the time, while men say it is the women group only 67\% (95\% CI [.51, .80]) of the time.

Building on Study 1's finding that participants do not even all agree on whether an object is a stereotype or not (and if it is, whether it is harmful), this analysis further shows that even when participants are in agreement that an object is a stereotype, they are not necessarily in agreement about why. The same holds true for whether a stereotype is harmful. One potential implication of this is considering whether different reasoning should lead to different bias mitigation. For example, if the reason an object is a stereotype is descriptive, then mitigation should aim to change the cognitive representations of people. To change these descriptive statistics, while we can work to alter the model outputs, we should also work to change society, the burden of which falls on a much larger group than just machine learning practitioners, e.g., policymakers. On the other hand, if particular stereotypes are deemed harmful because they are proscriptive and seem to restrict people from various avenues, we can consider ways to break free of gender norms, including by representing groups in ways that depart from those proscriptive norms.

\section{Study 3a: Errors that reinforce stereotypes may not lead to pragmatic harm}
Next, we test the pragmatic harm to users. Following the tripartite model of attitudes~\cite{allport1954nature, hamilton2014stereotypes, hilton1996stereotypes, katz1933racial, lippmann1922public, rosenberg1960cab} which differentiates between cognitive beliefs, emotions, and behaviors, we define pragmatic harm as negative changes in these three dimensions.

\subsection{Method}
To test pragmatic harm in stereotype-reinforcing errors, we conduct a between-subject survey experiment, using the stereotype-violating and neutral errors as control conditions (Figure~\ref{fig:error_types}). The cover story instructs participants to look at our synthesized search result page, imagining it is their personal phone photo album, and find a picture they had taken of someone they saw with a particular object. The search result page looks different for each randomized condition. We randomly assign participants to one of the three conditions (\textit{N} = 600): the stereotype-reinforcing condition exposes an image search result page with stereotype-reinforcing errors, e.g., false positive of \texttt{oven} on images of women; the stereotype-violating condition contains the same for stereotype-violating errors, e.g., false positive of \texttt{oven} on images of men; the stereotype-neutral condition contains neutral errors, e.g., false positive of \texttt{bowl} on images of women. The errors are actual misclassifications from a trained computer vision model to make them as closely aligned to realistic errors a user would encounter. We then measure participants’ cognitive beliefs, attitudes, and behaviors~\citep{fiske2002scm} to see if there are any changes because of such exposure. 
For \textit{cognitive beliefs}, we ask three sets of questions which span the spectrum of stereotype-specific to more generically about gendered beliefs. 
Concretely, we ask about: estimations of who uses ovens and bowls more between men and women; estimations of who tends to be the homemaker more between men and women; and perceived levels of warmth and competence~\cite{fiske2002scm} of women.
To assess \textit{attitude}, we ask two sets of questions. The first is about how participants feel about women in terms of four emotional components that are believed to mediate interactions between cognitive beliefs and behaviors: a) respect or admiration, b) pity or sympathy, c) disgust or sickening, and d) jealousy or envy~\cite{cuddy2007bias, fiske2002emotions, seger2017emotion}.
The second asks about sexist attitudes via a shortened scale focused on benevolent sexism ~\cite{glick1996asi, rollero2014shortened, glick2010hostility}.\footnote{We ask questions from the Ambivalent Sexism Inventory~\cite{glick1996asi} about benevolent sexism, as opposed to hostile sexism, because the latter is believed to suffer heavily from social desirability bias.} 
Finally, for \textit{behavioral} measures, we ask participants to undertake a realistic task they are liable to encounter which can cause harm: data labeling~\cite{miltenburg2016stereotype}. We chose this behavior measure because online participants are often the source of training labels in large-scale machine learning datasets. We ask participants to perform two common types of labeling on image data: tagging and captioning. In the tagging task, we ask participants to label the top three most relevant tags in an image which contains both the stereotype object (e.g., oven) and neutral object (e.g., bowl). We alter the perceived gender of the person to assess whether this changes what is tagged in the image. For the captioning task we show two people, one who looks masculine and another feminine, and swap whether there is a bowl or oven present in the image. This is to understand if the annotators will differently describe who is interacting with the object depending on whether it is stereotypically associated with women or not. All images are generated and/or manipulated by DALL-E 2. Additional details are in Appendix~\ref{app:methods}.
If stereotype-reinforcing errors have an influence on participants' cognitive representations, attitudes, and tagging or captioning behaviors, we should expect to see a statistically significant difference between participants who are exposed to search results with \texttt{oven}-women and those who are exposed to search results with \texttt{oven}-men or \texttt{bowl}-women. 

\subsection{Results}
\begin{figure*}[h]
    \centering
    \includegraphics[width=1\textwidth]{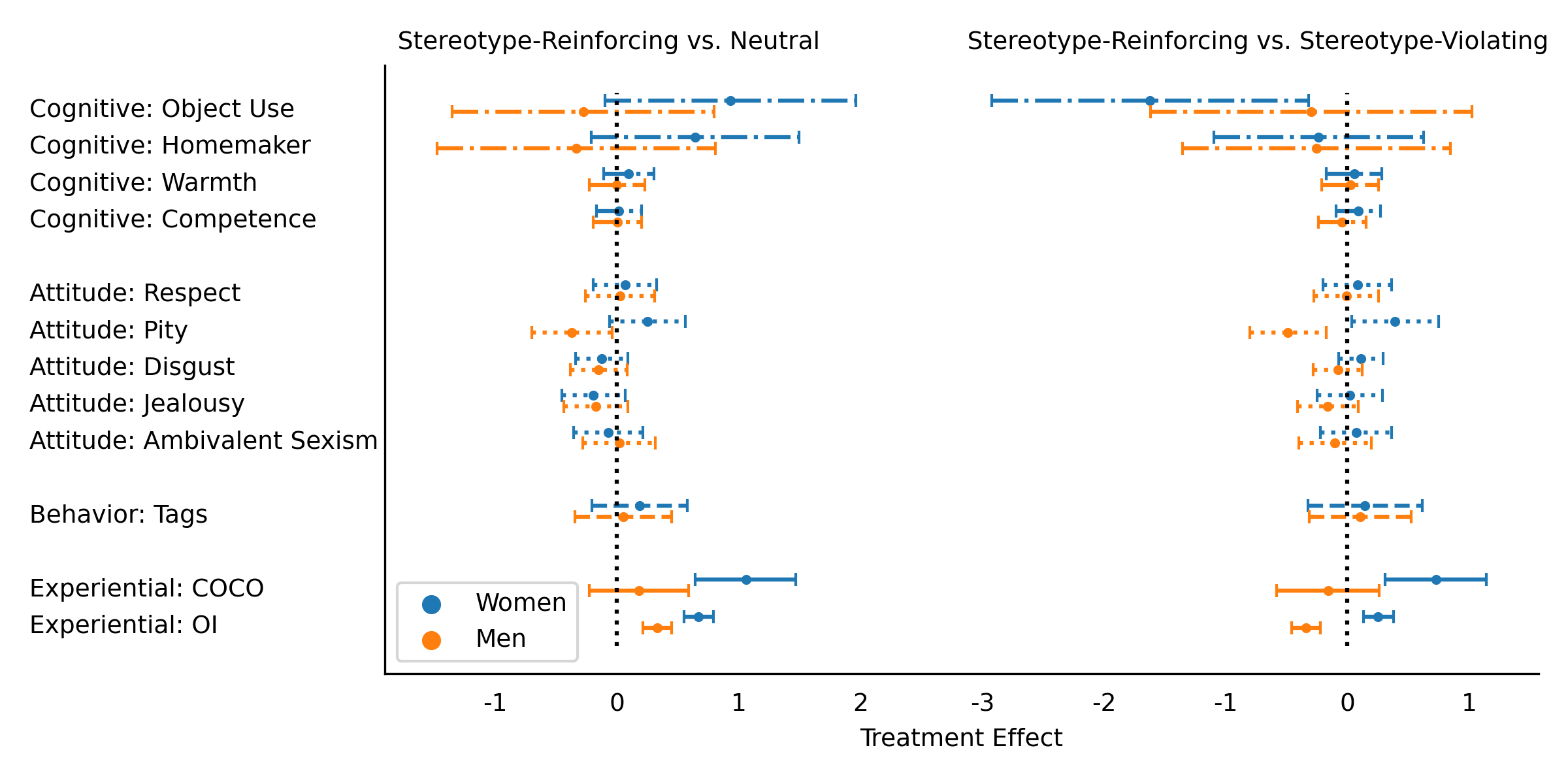}
    \caption{\textbf{Study 3, 4 Results.} The effect sizes and 95\% confidence intervals are reported for 10 of our 11 measures of pragmatic harm (for the behavior measure of captioning, we provide a descriptive analysis), experiential harm on COCO, and experiential harm on our larger dataset of OpenImages. Deviations from zero indicate that exposure to the stereotype-reinforcing stimulus resulted in our measured harm compared to exposure to the control condition.}
    \label{fig:quantfig}
\end{figure*}

Contrary to what we had originally hypothesized, after adjusting for multiple comparisons we do not find hypothesized statistically significant differences. 
We run an Ordinary-Least-Square (OLS) regression with the control condition coded as 0 and the experimental condition coded as 1, composite scores for beliefs, attitudes, and behaviors respectively as the dependent variables. Results are shown in Fig.~\ref{fig:quantfig}. In other words, participants' cognitive beliefs, attitudes, and behaviors towards gender groups, although stereotypical, remain stable before and after the treatment effect.

Prior work looking at a subset of these pragmatic harms has found very small effects in terms of cognitive belief changes about the representation of gendered occupations~\citep{kay2015search, metaxa2021image}.
Another line of work that finds a cognitive effect takes a different approach by studying novel occupations for which there are very few preconceived stereotypes~\citep{vlasceanu2022internet}. In our work, we focus on the activation of existing stereotypes, rather than the inducement of novel stereotypes.
Overall we find that the pragmatic harms are not measurable after exposure in the current survey experiment, which serves as a conservative test of harm, likely because the effects of these harms are too diffuse and long-term, impacted by all of the facets of society we encounter in our lives~\citep{noble2018oppression}. 

\section{Study 3b: Errors that reinforce stereotype lead to experiential harm}
Finally, we measure users’ experiential harm in response to different machine learning errors. This includes self-reports of negative affect based on the Positive and Negative Affect Schedule (PANAS; ~\citep{watson1988panas, crawford2004panas}). This is used as a measure of dignitary harms~\cite{katzman2023taxonomy}. In conceptualizing the experiential harm of machine learning errors which may seem individually minor, we draw a parallel to the concept of microaggressions. Microaggressions are ``small act[s] of insult or indignity, relating to a person’s membership in a socially oppressed group, which seems minor on its own but plays a part in significant systemic harm'' \citep{rini2020microaggression}. 
Just like how a machine learning model's classification error (e.g., of an \texttt{oven} on an image of a woman) may seem small on its own, and are ``easily interpretable as inadvertent errors rather than as malevolent actions,'' their negative effects on the target are real and should not be neglected~\citep{rini2020microaggression}.

\subsection{Method}
In terms of experiential harm, we design a within-subjects survey experiment ($N = 100$). We operationalize experiential harm by explicitly asking participants to rate how personally harmful they find different kinds of errors (which are stereotype-reinforcing, stereotype-violating, or neutral), on a scale from 0 (not at all) to 9 (extremely). This experience of error is analogous to situations where one reads in the news about the types of errors that artificial intelligence systems make~\citep{simonite2018comes}, notices such a pattern of errors themselves, or is informed by a friend. 

\subsection{Results}
In line with what we hypothesized, when we compare stereotype-reinforcing against neutral errors, an OLS regression shows participants rate stereotype-reinforcing errors to be more harmful than neutral ones ($b = .62$, 95\% \textit{CI} [.32, .91], $p<.001$). However, when disaggregating by gender this effect is only present among women participants (women: $b = 1.06$, 95\% \textit{CI} [.64, 1.47], $p<.001$; men: $b = .18$, 95\% \textit{CI} [-.23, .59], $p = .393$). When we use the stereotype-violating error as the control condition rather than the neutral error, we again find participants rate stereotype-reinforcing errors to be more harmful, though to a smaller degree, ($b = .28$, 95\% \textit{CI} [-.01, .58], $p = .062$), once again an effect only for women participants (women: $b = .73$, 95\% $CI$ [.31, 1.14], $p = .001$; men: $b = -.16$, 95\% $CI$ [-.58, .26], $p = .453$). See Fig.~\ref{fig:quantfig}.

In short, while we find little immediate evidence of pragmatic harms, we do find the existence of experiential harms resulting from stereotype-reinforcing errors, compared to both stereotype-violating and neutral errors. However, this pattern is present only among woman participants, and not men participants. 



\section{Study 4: Errors that violate stereotypes can be perceived as harmful too}

\subsection{Method}
We first test the generalizability of the previous findings by replicating the experiment and analysis with another popular dataset in object recognition tasks which is much larger: OpenImages~\citep{kuznetsova2020openimages}. 

Then, to test whether participants experience more experiential harm when they are exposed to stereotype-reinforcing (e.g., \texttt{skirt} on women), stereotype-violating (e.g., \texttt{skirt} on men), and neutral (e.g., \texttt{toothbrush} on women) errors, we use a similar procedure as in Study 3b. Rather than asking simply about ``personal harm'' as we did in Study 3b, here we draw from the Positive and Negative Affect Schedule (PANAS; ~\citep{watson1988panas, crawford2004panas}) and provide more details by asking about if they experience harm such as feeling upset, irritated, ashamed, or distressed. We conduct a within-subjects study and ask participants (\textit{N} = 300) to report their subjective experiences on a Likert scale from 0 to 9 for errors which vary in their relationship to a stereotype. The analysis uses a mixed-effects regression with experimental conditions as the independent variable, a composite score of experiential harm as the dependent variable, participants’ gender as the covariate variable, and error terms clustered at the individual level.

Then, we explore a new hypothesis that is not pre-registered about gender presentation-aligned objects, e.g., clothing, to dive deeper into the 40 stereotypical objects to understand why stereotype-violating errors are sometimes perceived to be more experientially harmful than stereotype-reinforcing ones. According to the gender trouble framework, costume (i.e., body and appearance) and script (i.e., behavior, traits, and preferences) are two aspects of gender performance, and reactions to androgynous or conventionally contradictory components can differ depending on which of the two it manifests in~\citep{butler1990gendertrouble, morgenroth2020gendertrouble, stern2017androgyny, goffman1959presentation, morgenroth2018gendertroublesocialpsych}. We thus hypothesize that conventionally contradictory costume objects may evoke more negative reactions compared to conventionally contradictory script objects~\citep{rudman2012vanguards}.
To test this hypothesis, we explore an additional independent variable we call ``wearable.'' We determined the value of this variable by manually marking 13 of the 40 stereotypical objects to be conventionally wearable by a person. These include objects like \texttt{football helmet} and \texttt{lipstick}, and exclude those like \texttt{truck} or \texttt{wine glass}. 

\subsection{Results}
OpenImages has 600 objects, annotated with perceived binary genders of people present in the image if applicable~\citep{schumann2021miap}. 
Following the same procedure as in the COCO dataset with new online participants (\textit{N} = 120), we find 249 of the 600 objects are marked as stereotypes by more than half of the participants, replicating the finding that not all objects are perceived as stereotypes. We then compile a list of 40 stereotypical objects (20 about men: e.g., \texttt{football}, \texttt{tool}; 20 about women: e.g., \texttt{doll}, \texttt{lipstick}), and 20 neutral objects (e.g., \texttt{balloon}, \texttt{goldfish}) for this study.

Replicating Study 3b, participants find stereotype-reinforcing errors to be more harmful than neutral ones ($b$ = .50, 95\% $CI$[.42, .59], $p<.001$). Again, this pattern is more pronounced among women participants ($b$ = .67, 95\% $CI$ [.55, .79], $p < .001$), with now a small effect among men participants ($b$ = .33, 95\% $CI$ [.21, .45], $p < .001$). Unlike Study 3b, we do not see differences in experiential harm between stereotype-reinforcing and stereotype-violating conditions ($b$ = -.04, 95\% $CI$ [-.13, .05], $p = .338$). The effect is canceled out by the opposite effects for women ($b$ = .25, 95\% $CI$ [.13, .38], $p <.001$) and men ($b$ = -.34, 95\% $CI$ [-.46, -.22], $p < .001$) participants. In other words, while women participants feel upset, irritated, ashamed, and distressed when they see stereotype-reinforcing errors (e.g., skirt on women), men participants feel that way when they see stereotype-violating errors (e.g., skirt on men). Results are in Fig.~\ref{fig:quantfig}.

To better understand this finding, with the ``wearable" distinction, we find that participants do rate stereotype-reinforcing errors to be more harmful than stereotype-violating ones (b = .23 95\% CI [.12, .34], $p < .001$), though again this effect exists in women participants (b = .49, 95\% CI [.34, .64], $p < .001$) rather than men participants (b = -.03, 95\% CI [-.18, .12], $p = .726$). 
Notably, for the interaction effect of a ``wearable'' object with the condition type, we find that wearable stereotype-violating errors have higher experiential harm than wearable stereotype-reinforcing errors ($b$=.80, 95\% $CI$ [.62, .99], $p<.001$), which is higher for men ($b$=.94, 95\% $CI$ [.67, 1.12], $p<.001$) than women participants ($b$=.69, 95\% $CI$ [.43, .94], $p<.001$). In other words, men participants find it more harmful than women participants when lipstick is misclassified on a man than on a woman.


In sum, stereotype-violating errors seem to cause harm too, possibly through different mechanisms. In addition to this result being a consequence of backlash effects~\citep{rudman2012backlash}, we raise two more possible mechanisms. First, it could be seen as an expression of precarious manhood; a concept that suggests manhood is precarious and needs continuous social validation such that threats to traditional masculinity can provoke anxiety in men, thus resulting in higher reports of harm~\citep{vandello2008precarious}. Second, these results may reflect elements of transphobia, which involves a negative reaction to the apparent incongruity between a person's perceived gender and a wearable gender presentation item~\citep{butler1990gendertrouble, morgenroth2020gendertrouble}. The divergent effect between men and women participants aligns with research indicating that transphobia is higher amongst cisgender men when judging transgender women due to the perceived threat to masculinity~\citep{makwana2018transphobia, nagoshi2019prejudices}. This analysis pushes us to reevaluate how we should think about reducing experiential harm, as it may encompass intolerances we do not wish to support.

\section{Discussion}

Taken together, our studies highlight the \textit{complexity} of machine learning errors. We offer three key contributions: (1) In contrast to prior assumptions which rely on a restricted set of stereotypes based on researcher intuitions, we present a much broader set of stereotypes that are defined by participants. (2) Compared to general notions of stereotypes or harms, we distinguish between machine learning errors that are stereotype-reinforcing, stereotype-violating, or stereotype-neutral, and formulate harm as pragmatic (changes in cognition, attitude, and behaviors) or experiential (self-report negative affect). (3) Our empirical experimental studies suggest that stereotype-reinforcing errors cause experiential, but not pragmatic, harm. Such experiential harm is unequally distributed, impacting more women participants than men. We speculate that long-term observational studies, compared to lab studies, are likely better suited to capture pragmatic harms~\citep{ford1997effects, fujioka1999examination, jenningswalstedt1980commercials}.

Formulating concrete notions of harm as we have done has implications beyond just machine learning: legal documents like the European AI Act are beginning to incorporate notions of psychological harm but lack definitions in which to ground regulation~\citep{bayefsky2016psychologicalharm, palka2023harm}. We also find stereotype-violating errors to be experientially harmful, especially when these errors pertain to wearable items associated with gender presentation. This effect is stronger for participants who identify as men compared to as women. This point warrants an especially nuanced discussion, as it questions whether we should take people's words at face value when they indicate something is personally harmful. To navigate this complexity, we can turn to the notions of epistemic injustice~\citep{fricker2009epistemic} and standpoint epistemology~\citep{wylie2003standpoint,fatima2020know, odowd2018microaggression}. If we interpret the negative reactions to misclassifications of stereotypically feminine clothing items on men as a manifestation of precarious manhood~\citep{vandello2008precarious} or transphobia~\citep{butler1990gendertrouble}, then we should down weight these concerns in designing mitigation approaches. Respecting people's experiential harms may not be as simple as accepting them at face value for direct measurement, but rather involves understanding which groups are likely to be harmed by what kinds of errors and why. This underscores the importance of work about diversity in annotators' perspectives~\citep{denton2021individual, waseem2016racist, davani2022disagreements, noble2012crowdsourcing, kairam2016crowds, dumitrache2018ambiguity}, and how much complexity is reduced by the use of discrete labels. 
Qualitative follow-up questions supplemented our annotations, where a lack of consensus is not a weakness or artifact to be averaged out, but rather a point for deeper inquiry on how to prioritize differential experiences.

Our findings call for reconsidering fairness measurement in supervised machine learning tasks. This involves considering how we can leverage human-driven insights to inform model training and evaluation~\citep{boykin2021interdisciplinary}. Traditionally, fairness evaluations tend to focus on stereotypes only in relation to occupations or traits. However our work expands this idea by showing that labels such as objects can also give rise to such harms.
Additionally, most prior work has only considered the implications of errors that reinforce stereotypes, which is relatively more intuitive to think of as harmful. However, both practically and normatively, it is important to understand the implications of stereotype-violating errors. Practically, strategies aimed at mitigating stereotype-reinforcing errors which act upon the target label will inevitably impact the occurrence of stereotype-violating errors as well. 
And normatively, there are also questions about whether stereotype-violating errors may even play a role in reducing stereotypical associations by counteracting them.

This finding that not only are certain labels more liable to cause harm than others, but that it matters for \textit{which} demographic group that label is misclassified, suggests that generic approaches like having a higher threshold for the classification of certain labels are insufficient. Instead, more nuanced fairness-through-awareness approaches~\citep{dwork2012awareness} will need to be taken. 
While adopting a cost-sensitive framework~\citep{kukar1998cost}
(e.g., different costs are associated with false positives and false negatives) is a simplified interpretation of our findings, it could be a starting point as one grapples with the questions of whose levels of harms we would prioritize reducing in a bias mitigation framework.
Our findings are limited by the methodological choices we made. For instance, we focused on gender stereotypes and do not know to what extent this finding generalizes to other groups such as race and age. Second, we recruited online participants in America who identify as men and women, missing other populations such as non-binary individuals or those from other cultural contexts. Given that stereotypes are culture-sensitive, and our work also shows that the harm perception is identity-sensitive, future work needs to study the interaction between participants' identity, culture, and harm perceptions. 
Third, the survey experiment does not capture harms beyond the two we measure (e.g., stereotype-threat~\citep{steele1996stereotypethreat, spencer2015threat}), nor the longitudinal effects of machine learning effects. Future work needs to capture not only the plurality of harms from machine learning errors but also how their effects over time.

Overall, drawing on research and methods from psychology, our work offers a rigorous empirical study connecting machine learning outputs to concrete harms by understanding the impact of stereotypical misclassifications. Rather than gesturing at harm as a justification for fairness measurement, we are very concrete in our analysis of the effects on people. Our finding that stereotype-reinforcing errors are experientially harmful for women underscores the importance for machine learning fairness interventions to be more rooted in social contexts, moving beyond objectives like just achieving equal prediction performance across groups. The diversity of responses we've presented, each influenced by participants' unique rationales, suggests the need for greater exploration of human psychological experiences in understanding how machine learning can cause harm.


\subsection*{Ethical considerations} As this is a study that involves human subjects, we sought and obtained IRB approval from our institution. Our study ran the risk of inducing negative feelings among our participants, given that we were investigating whether certain conditions caused subjective harm. That said, our study aimed to simulate conditions that participants likely encounter in their everyday lives already and thus tried to avoid exposing them to greater discomfort than they might experience normally. We also recruited only participants who identified as men or women, and asked questions about stereotypes related to binary gender, which may have caused its own kind of subjective harm to any non-binary participants that we did not enlist in the study.

\subsection*{Researcher positionality} 
All of us identify as cisgender and thus do not have the direct lived experience of the potentially transphobic harms we find in our work.

\subsection*{Adverse impact statement} While our study does not find statistically significant differences in beliefs, attitudes, and behaviors immediately following exposure to stereotype-associated machine learning errors, this does not mean that such errors do not give rise to cumulative, long-term pragmatic harms. There is a risk that our result will be misinterpreted to support the view that such errors are not harmful or worth addressing, or that such pragmatic harms are too difficult to meaningfully measure. Moreover, the experiential harms that we do detect might be dismissed as purely subjective and therefore relatively insignificant, despite these taking a significant emotional toll.  
Finally, we emphasize that our takeaway from this finding is not to take self-reports of harm always at face value, but rather to think critically about why people self-report harm and what harms they are actually experiencing.

\section*{Acknowledgments}
This material is based upon work supported by the National Science Foundation Graduate Research Fellowship to Angelina Wang. We are grateful to funding from the Data-Driven Social Science Initiative at Princeton University.
We thank Orly Bareket, Molly Crockett, Sunnie S. Y. Kim, Anne Kohlbrenner, Danaë Metaxa, Vikram V. Ramaswamy, Olga Russakovsky, Hanna Wallach, and members of the Visual AI Lab at Princeton, Fiske Lab at Princeton, and Perception and Judgment Lab at the University of Chicago for feedback.

\bibliographystyle{ACM-Reference-Format}
\bibliography{references}


\begin{thebibliography}{120}


\ifx \showCODEN    \undefined \def \showCODEN     #1{\unskip}     \fi
\ifx \showDOI      \undefined \def \showDOI       #1{#1}\fi
\ifx \showISBNx    \undefined \def \showISBNx     #1{\unskip}     \fi
\ifx \showISBNxiii \undefined \def \showISBNxiii  #1{\unskip}     \fi
\ifx \showISSN     \undefined \def \showISSN      #1{\unskip}     \fi
\ifx \showLCCN     \undefined \def \showLCCN      #1{\unskip}     \fi
\ifx \shownote     \undefined \def \shownote      #1{#1}          \fi
\ifx \showarticletitle \undefined \def \showarticletitle #1{#1}   \fi
\ifx \showURL      \undefined \def \showURL       {\relax}        \fi
\providecommand\bibfield[2]{#2}
\providecommand\bibinfo[2]{#2}
\providecommand\natexlab[1]{#1}
\providecommand\showeprint[2][]{arXiv:#2}

\bibitem[Abbasi et~al\mbox{.}(2019)]%
        {abbasi2019stereotyping}
\bibfield{author}{\bibinfo{person}{Mohsen Abbasi}, \bibinfo{person}{Sorelle~A.
  Friedler}, \bibinfo{person}{Carlos Scheidegger}, {and}
  \bibinfo{person}{Suresh Venkatasubramanian}.}
  \bibinfo{year}{2019}\natexlab{}.
\newblock \showarticletitle{Fairness in representation: quantifying
  stereotyping as a representational harm}.
\newblock \bibinfo{journal}{\emph{Siam International Conference on Data
  Mining}} (\bibinfo{year}{2019}).
\newblock


\bibitem[Allport et~al\mbox{.}(1954)]%
        {allport1954nature}
\bibfield{author}{\bibinfo{person}{Gordon~Willard Allport},
  \bibinfo{person}{Kenneth Clark}, {and} \bibinfo{person}{Thomas Pettigrew}.}
  \bibinfo{year}{1954}\natexlab{}.
\newblock \showarticletitle{The nature of prejudice}.
\newblock \bibinfo{journal}{\emph{Addison-wesley Reading, MA}}
  (\bibinfo{year}{1954}).
\newblock


\bibitem[Argyle et~al\mbox{.}(2023)]%
        {argyle2023many}
\bibfield{author}{\bibinfo{person}{Lisa~P. Argyle}, \bibinfo{person}{Ethan~C.
  Busby}, \bibinfo{person}{Nancy Fulda}, \bibinfo{person}{Joshua Gubler},
  \bibinfo{person}{Christopher Rytting}, {and} \bibinfo{person}{David
  Wingate}.} \bibinfo{year}{2023}\natexlab{}.
\newblock \showarticletitle{Out of One, Many: Using Language Models to Simulate
  Human Samples}.
\newblock \bibinfo{journal}{\emph{Political Analysis}} (\bibinfo{year}{2023}).
\newblock


\bibitem[Barlas et~al\mbox{.}(2021)]%
        {barlas2021see}
\bibfield{author}{\bibinfo{person}{Pinar Barlas}, \bibinfo{person}{Kyriakos
  Kyriakou}, \bibinfo{person}{Olivia Guest}, \bibinfo{person}{Styliani
  Kleanthous}, {and} \bibinfo{person}{Jahna Otterbacher}.}
  \bibinfo{year}{2021}\natexlab{}.
\newblock \showarticletitle{To "See" is to Stereotype: Image Tagging
  Algorithms, Gender Recognition, and the Accuracy-Fairness Trade-off}.
\newblock \bibinfo{journal}{\emph{Proceedings of the ACM on Human-Computer
  Interaction (CSCW)}} (\bibinfo{year}{2021}).
\newblock


\bibitem[Bayefsky(2016)]%
        {bayefsky2016psychologicalharm}
\bibfield{author}{\bibinfo{person}{Rachel Bayefsky}.}
  \bibinfo{year}{2016}\natexlab{}.
\newblock \showarticletitle{Psychological Harm and Constitutional Standing}.
\newblock \bibinfo{journal}{\emph{Brooklyn Law Review}} (\bibinfo{year}{2016}).
\newblock


\bibitem[Beeghly(2015)]%
        {beeghly2015stereotype}
\bibfield{author}{\bibinfo{person}{Erin Beeghly}.}
  \bibinfo{year}{2015}\natexlab{}.
\newblock \showarticletitle{What is a Stereotype? What is Stereotyping?}
\newblock \bibinfo{journal}{\emph{Hypatia}} (\bibinfo{year}{2015}).
\newblock


\bibitem[Beeghly(2025)]%
        {beeghly2025stereotype}
\bibfield{author}{\bibinfo{person}{Erin Beeghly}.}
  \bibinfo{year}{2025}\natexlab{}.
\newblock \showarticletitle{What's Wrong with Stereotyping?}
\newblock \bibinfo{journal}{\emph{Oxford University Press}}
  (\bibinfo{year}{2025}).
\newblock


\bibitem[Bhaskaran and Bhallamudi(2019)]%
        {bhaskaran2019secretaries}
\bibfield{author}{\bibinfo{person}{Jayadev Bhaskaran} {and}
  \bibinfo{person}{Isha Bhallamudi}.} \bibinfo{year}{2019}\natexlab{}.
\newblock \showarticletitle{Good Secretaries, Bad Truck Drivers? Occupational
  Gender Stereotypes in Sentiment Analysis}.
\newblock \bibinfo{journal}{\emph{Proceedings of the First Workshop on Gender
  Bias in Natural Language Processing}} (\bibinfo{year}{2019}).
\newblock


\bibitem[Bianchi et~al\mbox{.}(2023)]%
        {bianchi2023generation}
\bibfield{author}{\bibinfo{person}{Federico Bianchi},
  \bibinfo{person}{Pratyusha Kalluri}, \bibinfo{person}{Esin Durmus},
  \bibinfo{person}{Faisal Ladhak}, \bibinfo{person}{Myra Cheng},
  \bibinfo{person}{Debora Nozza}, \bibinfo{person}{Tatsunori~B Hashimoto},
  \bibinfo{person}{Dan~Saul Jurafsky}, \bibinfo{person}{James Zou}, {and}
  \bibinfo{person}{Aylin Caliskan}.} \bibinfo{year}{2023}\natexlab{}.
\newblock \showarticletitle{Easily Accessible Text-to-Image Generation
  Amplifies Demographic Stereotypes at Large Scale}.
\newblock \bibinfo{journal}{\emph{ACM Conference on Fairness, Accountability,
  and Transparency (FAccT)}} (\bibinfo{year}{2023}).
\newblock


\bibitem[Blodgett et~al\mbox{.}(2021)]%
        {blodgett2021salmon}
\bibfield{author}{\bibinfo{person}{Su~Lin Blodgett}, \bibinfo{person}{Gilsinia
  Lopez}, \bibinfo{person}{Alexandra Olteanu}, \bibinfo{person}{Robert Sim},
  {and} \bibinfo{person}{Hanna Wallach}.} \bibinfo{year}{2021}\natexlab{}.
\newblock \showarticletitle{Stereotyping Norwegian Salmon: An Inventory of
  Pitfalls in Fairness Benchmark Datasets}.
\newblock \bibinfo{journal}{\emph{Proceedings of the 59th Annual Meeting of the
  Association for Computational Linguistics and the 11th International Joint
  Conference on Natural Language Processing}} (\bibinfo{year}{2021}).
\newblock


\bibitem[Bolukbasi et~al\mbox{.}(2016)]%
        {bolukbasi2016embeddings}
\bibfield{author}{\bibinfo{person}{Tolga Bolukbasi}, \bibinfo{person}{Kai-Wei
  Chang}, \bibinfo{person}{James Zou}, \bibinfo{person}{Venkatesh Saligrama},
  {and} \bibinfo{person}{Adam Kalai}.} \bibinfo{year}{2016}\natexlab{}.
\newblock \showarticletitle{Man is to Computer Programmer as Woman is to
  Homemaker? Debiasing Word Embeddings}.
\newblock \bibinfo{journal}{\emph{Conference on Neural Information Processing
  Systems (NeurIPS)}} (\bibinfo{year}{2016}).
\newblock


\bibitem[Boykin et~al\mbox{.}(2021)]%
        {boykin2021interdisciplinary}
\bibfield{author}{\bibinfo{person}{C.~Malik Boykin}, \bibinfo{person}{Sophia~T.
  Dasch}, \bibinfo{person}{Vincent~Rice Jr.}, \bibinfo{person}{Venkat~R.
  Lakshminarayanan}, \bibinfo{person}{Taiwo~A. Togun}, {and}
  \bibinfo{person}{Sarah~M. Brown}.} \bibinfo{year}{2021}\natexlab{}.
\newblock \showarticletitle{Opportunities for a More Interdisciplinary Approach
  to Measuring Perceptions of Fairness in Machine Learning}.
\newblock \bibinfo{journal}{\emph{Equity and Access in Algorithms, Mechanisms,
  and Optimization (EAAMO)}} (\bibinfo{year}{2021}).
\newblock


\bibitem[Burgess and Borgida(1999)]%
        {burgess1999descriptiveprescriptive}
\bibfield{author}{\bibinfo{person}{Diana Burgess} {and} \bibinfo{person}{Eugene
  Borgida}.} \bibinfo{year}{1999}\natexlab{}.
\newblock \showarticletitle{Who women are, who women should be: descriptive and
  prescriptive gender stereotyping in sex discrimination}.
\newblock \bibinfo{journal}{\emph{Psychology, Public Policy, and Law}}
  \bibinfo{volume}{5} (\bibinfo{year}{1999}).
\newblock
Issue 3.


\bibitem[Butler(1990)]%
        {butler1990gendertrouble}
\bibfield{author}{\bibinfo{person}{Judith Butler}.}
  \bibinfo{year}{1990}\natexlab{}.
\newblock \showarticletitle{Gender Trouble: Feminism and the Subversion of
  Identity}.
\newblock \bibinfo{journal}{\emph{Routledge}} (\bibinfo{year}{1990}).
\newblock


\bibitem[Caliskan et~al\mbox{.}(2017)]%
        {caliskan2017semantics}
\bibfield{author}{\bibinfo{person}{Aylin Caliskan}, \bibinfo{person}{Joanna~J.
  Bryson}, {and} \bibinfo{person}{Arvind Narayanan}.}
  \bibinfo{year}{2017}\natexlab{}.
\newblock \showarticletitle{Semantics derived automatically from language
  corpora contain human-like biases}.
\newblock \bibinfo{journal}{\emph{Science}} (\bibinfo{year}{2017}).
\newblock


\bibitem[Cao et~al\mbox{.}(2022)]%
        {cao2022theory}
\bibfield{author}{\bibinfo{person}{Yang Cao}, \bibinfo{person}{Anna Sotnikova},
  \bibinfo{person}{Hal~Daumé III}, \bibinfo{person}{Rachel Rudinger}, {and}
  \bibinfo{person}{Linda Zou}.} \bibinfo{year}{2022}\natexlab{}.
\newblock \showarticletitle{Theory-Grounded Measurement of U.S. Social
  Stereotypes in English Language Models}.
\newblock \bibinfo{journal}{\emph{Conference of the North American Chapter of
  the Association for Computational Linguistics: Human Language Technologies}}
  (\bibinfo{year}{2022}).
\newblock


\bibitem[Chien and Danks(2024)]%
        {chien2024behondbehaviorist}
\bibfield{author}{\bibinfo{person}{Jennifer Chien} {and} \bibinfo{person}{David
  Danks}.} \bibinfo{year}{2024}\natexlab{}.
\newblock \showarticletitle{Beyond Behaviorist Representational Harms: A Plan
  for Measurement and Mitigation}.
\newblock \bibinfo{journal}{\emph{ACM Conference on Fairness, Accountability,
  and Transparency (FAccT)}} (\bibinfo{year}{2024}).
\newblock


\bibitem[Crawford and Henry(2004)]%
        {crawford2004panas}
\bibfield{author}{\bibinfo{person}{John~R. Crawford} {and}
  \bibinfo{person}{Julie~D. Henry}.} \bibinfo{year}{2004}\natexlab{}.
\newblock \showarticletitle{The positive and negative affect schedule (PANAS):
  construct validity, measurement properties and normative data in a large
  non-clinical sample}.
\newblock \bibinfo{journal}{\emph{British Journal of Clinical Psychology}}
  (\bibinfo{year}{2004}).
\newblock


\bibitem[Cuddy et~al\mbox{.}(2007)]%
        {cuddy2007bias}
\bibfield{author}{\bibinfo{person}{Amy J.~C. Cuddy}, \bibinfo{person}{Susan~T.
  Fiske}, {and} \bibinfo{person}{Peter Glick}.}
  \bibinfo{year}{2007}\natexlab{}.
\newblock \showarticletitle{The {BIAS} map: behaviors from intergroup affect
  and stereotypes}.
\newblock \bibinfo{journal}{\emph{Journal of Personality and Social
  Psychology}}  \bibinfo{volume}{92} (\bibinfo{year}{2007}).
\newblock
Issue 4.


\bibitem[Czarnowska et~al\mbox{.}(2021)]%
        {czarnowska2021socialbiasnlp}
\bibfield{author}{\bibinfo{person}{Paula Czarnowska}, \bibinfo{person}{Yogarshi
  Vyas}, {and} \bibinfo{person}{Kashif Shah}.} \bibinfo{year}{2021}\natexlab{}.
\newblock \showarticletitle{Quantifying Social Biases in NLP: A Generalization
  and Empirical Comparison of Extrinsic Fairness Metrics}.
\newblock \bibinfo{journal}{\emph{Transactions of the Association for
  Computational Linguistics}} (\bibinfo{year}{2021}).
\newblock


\bibitem[Davani et~al\mbox{.}(2022)]%
        {davani2022disagreements}
\bibfield{author}{\bibinfo{person}{Aida~Mostafazadeh Davani},
  \bibinfo{person}{Mark Díaz}, {and} \bibinfo{person}{Vinodkumar
  Prabhakaran}.} \bibinfo{year}{2022}\natexlab{}.
\newblock \showarticletitle{Dealing with Disagreements: Looking Beyond the
  Majority Vote in Subjective Annotations}.
\newblock \bibinfo{journal}{\emph{Transactions of the Association for
  Computational Linguistics}} (\bibinfo{year}{2022}).
\newblock


\bibitem[Denton et~al\mbox{.}(2021)]%
        {denton2021individual}
\bibfield{author}{\bibinfo{person}{Emily Denton}, \bibinfo{person}{Mark Díaz},
  \bibinfo{person}{Ian Kivlichan}, \bibinfo{person}{Vinodkumar Prabhakaran},
  {and} \bibinfo{person}{Rachel Rosen}.} \bibinfo{year}{2021}\natexlab{}.
\newblock \showarticletitle{Whose Ground Truth? Accounting for Individual and
  Collective Identities Underlying Dataset Annotation}.
\newblock \bibinfo{journal}{\emph{NeurIPS 2021 Workshop on Data-Centric AI}}
  (\bibinfo{year}{2021}).
\newblock


\bibitem[Dev et~al\mbox{.}(2023)]%
        {dev2023stereotypeindia}
\bibfield{author}{\bibinfo{person}{Sunipa Dev}, \bibinfo{person}{Jaya Goyal},
  \bibinfo{person}{Dinesh Tewari}, \bibinfo{person}{Shachi Dave}, {and}
  \bibinfo{person}{Vinodkumar Prabhakaran}.} \bibinfo{year}{2023}\natexlab{}.
\newblock \showarticletitle{Building Socio-culturally Inclusive Stereotype
  Resources with Community Engagement}.
\newblock \bibinfo{journal}{\emph{Advances in Neural Information Processing
  Systems (NeurIPS) Datasets and Benchmarks Track}} (\bibinfo{year}{2023}).
\newblock


\bibitem[Dev et~al\mbox{.}(2020)]%
        {dev2020embeddings}
\bibfield{author}{\bibinfo{person}{Sunipa Dev}, \bibinfo{person}{Tao Li},
  \bibinfo{person}{Jeff Phillips}, {and} \bibinfo{person}{Vivek Srikumar}.}
  \bibinfo{year}{2020}\natexlab{}.
\newblock \showarticletitle{On Measuring and Mitigating Biased Inferences of
  Word Embeddings}.
\newblock \bibinfo{journal}{\emph{AAAI Technical Track: Natural Language
  Processing}} (\bibinfo{year}{2020}).
\newblock


\bibitem[Dev and Phillips(2019)]%
        {dev2019vector}
\bibfield{author}{\bibinfo{person}{Sunipa Dev} {and} \bibinfo{person}{Jeff
  Phillips}.} \bibinfo{year}{2019}\natexlab{}.
\newblock \showarticletitle{Attenuating Bias in Word Vectors}.
\newblock \bibinfo{journal}{\emph{International Conference on Artificial
  Intelligence and Statistics}} (\bibinfo{year}{2019}).
\newblock


\bibitem[Dev et~al\mbox{.}(2022)]%
        {dev2022biasnlp}
\bibfield{author}{\bibinfo{person}{Sunipa Dev}, \bibinfo{person}{Emily Sheng},
  \bibinfo{person}{Jieyu Zhao}, \bibinfo{person}{Aubrie Amstutz},
  \bibinfo{person}{Jiao Sun}, \bibinfo{person}{Yu Hou}, \bibinfo{person}{Mattie
  Sanseverino}, \bibinfo{person}{Jiin Kim}, \bibinfo{person}{Akihiro Nishi},
  \bibinfo{person}{Nanyun Peng}, {and} \bibinfo{person}{Kai-Wei Chang}.}
  \bibinfo{year}{2022}\natexlab{}.
\newblock \showarticletitle{On Measures of Biases and Harms in NLP}.
\newblock \bibinfo{journal}{\emph{Findings of the Association for Computational
  Linguistics: AACL-IJCNLP}} (\bibinfo{year}{2022}).
\newblock


\bibitem[Devine and Elliot(1995)]%
        {devine1995trilogy}
\bibfield{author}{\bibinfo{person}{Patricia~G. Devine} {and}
  \bibinfo{person}{Andrew~J. Elliot}.} \bibinfo{year}{1995}\natexlab{}.
\newblock \showarticletitle{Are Racial Stereotypes Really Fading? The Princeton
  Trilogy Revisited}.
\newblock \bibinfo{journal}{\emph{Personality and Social Psychology Bulletin}}
  \bibinfo{volume}{21} (\bibinfo{year}{1995}).
\newblock
Issue 11.


\bibitem[Devlin et~al\mbox{.}(2019)]%
        {devlin2019bert}
\bibfield{author}{\bibinfo{person}{Jacob Devlin}, \bibinfo{person}{Ming-Wei
  Chang}, \bibinfo{person}{Kenton Lee}, {and} \bibinfo{person}{Kristina
  Toutanova}.} \bibinfo{year}{2019}\natexlab{}.
\newblock \showarticletitle{{BERT}: Pre-training of Deep Bidirectional
  Transformers for Language Understanding}.
\newblock \bibinfo{journal}{\emph{Proceedings of NAACL-HLT}}
  (\bibinfo{year}{2019}).
\newblock


\bibitem[Diaz(2008)]%
        {diaz2008goggles}
\bibfield{author}{\bibinfo{person}{Alejandro~M. Diaz}.}
  \bibinfo{year}{2008}\natexlab{}.
\newblock \showarticletitle{Through the Google Goggles: Sociopolitical Bias in
  Search Engine Design}.
\newblock \bibinfo{journal}{\emph{Information Science and Knowledge
  Management}} (\bibinfo{year}{2008}).
\newblock


\bibitem[Dosovitskiy et~al\mbox{.}(2021)]%
        {dosovitskiy2021vit}
\bibfield{author}{\bibinfo{person}{Alexey Dosovitskiy}, \bibinfo{person}{Lucas
  Beyer}, \bibinfo{person}{Alexander Kolesnikov}, \bibinfo{person}{Dirk
  Weissenborn}, \bibinfo{person}{Xiaohua Zhai}, \bibinfo{person}{Thomas
  Unterthiner}, \bibinfo{person}{Mostafa Dehghani}, \bibinfo{person}{Matthias
  Minderer}, \bibinfo{person}{Georg Heigold}, \bibinfo{person}{Sylvain Gelly},
  \bibinfo{person}{Jakob Uszkoreit}, {and} \bibinfo{person}{Neil Houlsby}.}
  \bibinfo{year}{2021}\natexlab{}.
\newblock \showarticletitle{An Image is Worth 16x16 Words: Transformers for
  Image Recognition at Scale}.
\newblock \bibinfo{journal}{\emph{International Conference on Learning
  Representations (ICLR)}} (\bibinfo{year}{2021}).
\newblock


\bibitem[Dumitrache et~al\mbox{.}(2018)]%
        {dumitrache2018ambiguity}
\bibfield{author}{\bibinfo{person}{Anca Dumitrache}, \bibinfo{person}{Lora
  Aroyo}, {and} \bibinfo{person}{Chris Welty}.}
  \bibinfo{year}{2018}\natexlab{}.
\newblock \showarticletitle{Capturing Ambiguity in Crowdsourcing Frame
  Disambiguation}.
\newblock \bibinfo{journal}{\emph{AAAI Conference on Human Computation and
  Crowdsourcing (HCOMP)}} (\bibinfo{year}{2018}).
\newblock


\bibitem[Dwork et~al\mbox{.}(2012)]%
        {dwork2012awareness}
\bibfield{author}{\bibinfo{person}{Cynthia Dwork}, \bibinfo{person}{Moritz
  Hardt}, \bibinfo{person}{Toniann Pitassi}, \bibinfo{person}{Omer Reingold},
  {and} \bibinfo{person}{Rich Zemel}.} \bibinfo{year}{2012}\natexlab{}.
\newblock \showarticletitle{Fairness Through Awareness}.
\newblock \bibinfo{journal}{\emph{Proceedings of the 3rd Innovations in
  Theoretical Computer Science Conference}} (\bibinfo{year}{2012}).
\newblock


\bibitem[Eagly(1987)]%
        {eagly1987socialrole}
\bibfield{author}{\bibinfo{person}{Alice~H. Eagly}.}
  \bibinfo{year}{1987}\natexlab{}.
\newblock \showarticletitle{Sex differences in social behavior: A social-role
  interpretation}.
\newblock \bibinfo{journal}{\emph{Lawrence Erlbaum Associates, Inc.}}
  (\bibinfo{year}{1987}).
\newblock


\bibitem[Eagly et~al\mbox{.}(2020)]%
        {eagly2020temporal}
\bibfield{author}{\bibinfo{person}{Alice~H. Eagly}, \bibinfo{person}{Christa
  Nater}, \bibinfo{person}{David~I. Miller}, \bibinfo{person}{Michèle
  Kaufmann}, {and} \bibinfo{person}{Sabine Sczesny}.}
  \bibinfo{year}{2020}\natexlab{}.
\newblock \showarticletitle{Gender stereotypes have changed: A cross-temporal
  meta-analysis of U.S. public opinion polls from 1946 to 2018}.
\newblock \bibinfo{journal}{\emph{American Psychologist}}
  (\bibinfo{year}{2020}).
\newblock


\bibitem[Ellemers et~al\mbox{.}(2018)]%
        {ellemers2018gender}
\bibfield{author}{\bibinfo{person}{Naomi Ellemers} {et~al\mbox{.}}}
  \bibinfo{year}{2018}\natexlab{}.
\newblock \showarticletitle{Gender stereotypes}.
\newblock \bibinfo{journal}{\emph{Annual review of psychology}}
  \bibinfo{volume}{69} (\bibinfo{year}{2018}), \bibinfo{pages}{275--298}.
\newblock


\bibitem[Fatima(2020)]%
        {fatima2020know}
\bibfield{author}{\bibinfo{person}{Saba Fatima}.}
  \bibinfo{year}{2020}\natexlab{}.
\newblock \showarticletitle{I Know What Happened to Me: The Epistemic Harms
  of Microaggression}.
\newblock \bibinfo{journal}{\emph{Microaggressions and Philosophy}}
  (\bibinfo{year}{2020}).
\newblock


\bibitem[Fiske et~al\mbox{.}(2002a)]%
        {fiske2002emotions}
\bibfield{author}{\bibinfo{person}{Susan~T. Fiske}, \bibinfo{person}{Amy J.~C.
  Cuddy}, {and} \bibinfo{person}{Peter Glick}.}
  \bibinfo{year}{2002}\natexlab{a}.
\newblock \showarticletitle{Emotions Up and Down: Intergroup Emotions Result
  from Status and Competition}.
\newblock \bibinfo{journal}{\emph{Prejudice to Intergroup Emotions:
  Differentiated Reactions to Social Groups}} (\bibinfo{year}{2002}).
\newblock


\bibitem[Fiske et~al\mbox{.}(2002b)]%
        {fiske2002scm}
\bibfield{author}{\bibinfo{person}{Susan~T. Fiske}, \bibinfo{person}{Amy J.~C.
  Cuddy}, \bibinfo{person}{Peter Glick}, {and} \bibinfo{person}{Jun Xu}.}
  \bibinfo{year}{2002}\natexlab{b}.
\newblock \showarticletitle{A model of (often mixed) stereotype content:
  Competence and warmth respectively follow from perceived status and
  competition}.
\newblock \bibinfo{journal}{\emph{Journal of Personality and Social
  Psychology}}  \bibinfo{volume}{82} (\bibinfo{year}{2002}).
\newblock
Issue 6.


\bibitem[Ford(1997)]%
        {ford1997effects}
\bibfield{author}{\bibinfo{person}{Thomas~E. Ford}.}
  \bibinfo{year}{1997}\natexlab{}.
\newblock \showarticletitle{Effects of Stereotypical Television Portrayals of
  African-Americans on Person Perception}.
\newblock \bibinfo{journal}{\emph{Social Psychology Quarterly}}
  (\bibinfo{year}{1997}).
\newblock


\bibitem[Fricker(2009)]%
        {fricker2009epistemic}
\bibfield{author}{\bibinfo{person}{Miranda Fricker}.}
  \bibinfo{year}{2009}\natexlab{}.
\newblock \showarticletitle{Epistemic Injustice: Power and the Ethics of
  Knowing}.
\newblock \bibinfo{journal}{\emph{Oxford University Press}}
  (\bibinfo{year}{2009}).
\newblock


\bibitem[Fujioka(1999)]%
        {fujioka1999examination}
\bibfield{author}{\bibinfo{person}{Yuki Fujioka}.}
  \bibinfo{year}{1999}\natexlab{}.
\newblock \showarticletitle{Television Portrayals and African-American
  Stereotypes: Examination of Television Effects when Direct Contact is
  Lacking}.
\newblock \bibinfo{journal}{\emph{Journalism and Mass Communication Quarterly}}
  (\bibinfo{year}{1999}).
\newblock


\bibitem[Gallegos et~al\mbox{.}(2023)]%
        {gallegos2023survey}
\bibfield{author}{\bibinfo{person}{Isabel~O. Gallegos},
  \bibinfo{person}{Ryan~A. Rossi}, \bibinfo{person}{Joe Barrow},
  \bibinfo{person}{Md~Mehrab Tanjim}, \bibinfo{person}{Sungchul Kim},
  \bibinfo{person}{Franck Dernoncourt}, \bibinfo{person}{Tong Yu},
  \bibinfo{person}{Ruiyi Zhang}, {and} \bibinfo{person}{Nesreen~K. Ahmed}.}
  \bibinfo{year}{2023}\natexlab{}.
\newblock \showarticletitle{Bias and Fairness in Large Language Models: A
  Survey}.
\newblock \bibinfo{journal}{\emph{arXiv:2309.00770}} (\bibinfo{year}{2023}).
\newblock


\bibitem[Garg et~al\mbox{.}(2018)]%
        {garg2018quantify}
\bibfield{author}{\bibinfo{person}{Nikhil Garg}, \bibinfo{person}{Londa
  Schiebinger}, \bibinfo{person}{Dan Jurafsky}, {and} \bibinfo{person}{James
  Zhou}.} \bibinfo{year}{2018}\natexlab{}.
\newblock \showarticletitle{Word embeddings quantify 100 years of gender and
  ethnic stereotypes}.
\newblock \bibinfo{journal}{\emph{Proceedings of the National Academy of
  Sciences of the United States of America (PNAS)}} (\bibinfo{year}{2018}).
\newblock


\bibitem[Glick and Fiske(1996)]%
        {glick1996asi}
\bibfield{author}{\bibinfo{person}{Peter Glick} {and} \bibinfo{person}{Susan~T.
  Fiske}.} \bibinfo{year}{1996}\natexlab{}.
\newblock \showarticletitle{The Ambivalent Sexism Inventory: Differentiating
  hostile and benevolent sexism}.
\newblock \bibinfo{journal}{\emph{Journal of Personality and Social
  Psychology}}  \bibinfo{volume}{70} (\bibinfo{year}{1996}).
\newblock
Issue 3.


\bibitem[Glick and Whitehead(2010)]%
        {glick2010hostility}
\bibfield{author}{\bibinfo{person}{Peter Glick} {and} \bibinfo{person}{Jessica
  Whitehead}.} \bibinfo{year}{2010}\natexlab{}.
\newblock \showarticletitle{Hostility Toward Men and the Perceived Stability of
  Male Dominance}.
\newblock \bibinfo{journal}{\emph{Social Psychology}}  \bibinfo{volume}{41}
  (\bibinfo{year}{2010}).
\newblock
Issue 3.


\bibitem[Goffman(1959)]%
        {goffman1959presentation}
\bibfield{author}{\bibinfo{person}{Erving Goffman}.}
  \bibinfo{year}{1959}\natexlab{}.
\newblock \showarticletitle{The Presentation of Self in Everyday Life}.
\newblock \bibinfo{journal}{\emph{Doubleday}} (\bibinfo{year}{1959}).
\newblock


\bibitem[Graves(1999)]%
        {graves1999television}
\bibfield{author}{\bibinfo{person}{Sherryl~Browne Graves}.}
  \bibinfo{year}{1999}\natexlab{}.
\newblock \showarticletitle{Television and prejudice reduction: When does
  television as a vicarious experience make a difference?m}.
\newblock \bibinfo{journal}{\emph{Journal of Social Issues}}
  \bibinfo{volume}{55} (\bibinfo{year}{1999}).
\newblock
Issue 4.


\bibitem[Greenwald et~al\mbox{.}(1998)]%
        {greenwald1998iat}
\bibfield{author}{\bibinfo{person}{Anthony~G. Greenwald},
  \bibinfo{person}{Debbie~E. McGhee}, {and} \bibinfo{person}{Jordan L.~K.
  Schwartz}.} \bibinfo{year}{1998}\natexlab{}.
\newblock \showarticletitle{Measuring individual differences in implicit
  cognition: the implicit association test}.
\newblock \bibinfo{journal}{\emph{Journal of Personality and Social
  Psychology}} (\bibinfo{year}{1998}).
\newblock


\bibitem[Grgić-Hlača et~al\mbox{.}(2018)]%
        {grgichlaca2018perception}
\bibfield{author}{\bibinfo{person}{Nina Grgić-Hlača},
  \bibinfo{person}{Elissa~M. Redmiles}, \bibinfo{person}{Krishna~P. Gummadi},
  {and} \bibinfo{person}{Adrian Weller}.} \bibinfo{year}{2018}\natexlab{}.
\newblock \showarticletitle{Human Perceptions of Fairness in Algorithmic
  Decision Making: A Case Study of Criminal Risk Prediction}.
\newblock \bibinfo{journal}{\emph{Proceedings of the Web Conference (WWW)}}
  (\bibinfo{year}{2018}).
\newblock


\bibitem[Guilbeault et~al\mbox{.}(2024)]%
        {guilbeault2024amplify}
\bibfield{author}{\bibinfo{person}{Douglas Guilbeault},
  \bibinfo{person}{Solène Delecourt}, \bibinfo{person}{Tasker Hull},
  \bibinfo{person}{Bhargav~Srinivasa Desikan}, \bibinfo{person}{Mark Chu},
  {and} \bibinfo{person}{Ethan Nadler}.} \bibinfo{year}{2024}\natexlab{}.
\newblock \showarticletitle{Online images amplify gender bias}.
\newblock \bibinfo{journal}{\emph{Nature}} (\bibinfo{year}{2024}).
\newblock


\bibitem[Hall et~al\mbox{.}(2022)]%
        {hall2022biasamp}
\bibfield{author}{\bibinfo{person}{Melissa Hall}, \bibinfo{person}{Laurens
  van~der Maaten}, \bibinfo{person}{Laura Gustafson}, \bibinfo{person}{Maxwell
  Jones}, {and} \bibinfo{person}{Aaron Adcock}.}
  \bibinfo{year}{2022}\natexlab{}.
\newblock \showarticletitle{A Systematic Study of Bias Amplification}.
\newblock \bibinfo{journal}{\emph{arXiv:2201.11706}} (\bibinfo{year}{2022}).
\newblock


\bibitem[Hamilton and Sherman(2014)]%
        {hamilton2014stereotypes}
\bibfield{author}{\bibinfo{person}{David~L Hamilton} {and}
  \bibinfo{person}{Jeffrey~W Sherman}.} \bibinfo{year}{2014}\natexlab{}.
\newblock \showarticletitle{Stereotypes}.
\newblock In \bibinfo{booktitle}{\emph{Handbook of social cognition}}.
  \bibinfo{publisher}{Psychology Press}, \bibinfo{pages}{17--84}.
\newblock


\bibitem[Hentschel et~al\mbox{.}(2019)]%
        {hentschel2019dimensions}
\bibfield{author}{\bibinfo{person}{Tanja Hentschel},
  \bibinfo{person}{Madeline~E. Heilman}, {and} \bibinfo{person}{Claudia~V.
  Peus}.} \bibinfo{year}{2019}\natexlab{}.
\newblock \showarticletitle{The Multiple Dimensions of Gender Stereotypes: A
  Current Look at Men’s and Women’s Characterizations of Others and
  Themselves}.
\newblock \bibinfo{journal}{\emph{Frontiers in Psychology}}
  (\bibinfo{year}{2019}).
\newblock


\bibitem[Hilton and Von~Hippel(1996)]%
        {hilton1996stereotypes}
\bibfield{author}{\bibinfo{person}{James~L Hilton} {and}
  \bibinfo{person}{William Von~Hippel}.} \bibinfo{year}{1996}\natexlab{}.
\newblock \showarticletitle{Stereotypes}.
\newblock \bibinfo{journal}{\emph{Annual review of psychology}}
  \bibinfo{volume}{47}, \bibinfo{number}{1} (\bibinfo{year}{1996}),
  \bibinfo{pages}{237--271}.
\newblock


\bibitem[Hämäläinen et~al\mbox{.}(2023)]%
        {hamalainen2023synthetic}
\bibfield{author}{\bibinfo{person}{Perttu Hämäläinen},
  \bibinfo{person}{Mikke Tavast}, {and} \bibinfo{person}{Anton Kunnari}.}
  \bibinfo{year}{2023}\natexlab{}.
\newblock \showarticletitle{Evaluating Large Language Models in Generating
  Synthetic HCI Research Data: a Case Study}.
\newblock \bibinfo{journal}{\emph{Conference on Human Factors in Computing
  Systems (CHI)}} (\bibinfo{year}{2023}).
\newblock


\bibitem[Introna and Nissenbaum(2000)]%
        {introna2000web}
\bibfield{author}{\bibinfo{person}{Lucas~D. Introna} {and}
  \bibinfo{person}{Helen Nissenbaum}.} \bibinfo{year}{2000}\natexlab{}.
\newblock \showarticletitle{Shaping the Web: Why the Politics of Search Engines
  Matters}.
\newblock \bibinfo{journal}{\emph{The Information Society}}
  (\bibinfo{year}{2000}).
\newblock


\bibitem[Jennings-Walstedt et~al\mbox{.}(1980)]%
        {jenningswalstedt1980commercials}
\bibfield{author}{\bibinfo{person}{Joyce Jennings-Walstedt},
  \bibinfo{person}{Florence~L. Geis}, {and} \bibinfo{person}{Virginia Brown}.}
  \bibinfo{year}{1980}\natexlab{}.
\newblock \showarticletitle{Influence of television commercials on women's
  self-confidence and independent judgment}.
\newblock \bibinfo{journal}{\emph{Journal of Personality and Social
  Psychology}} (\bibinfo{year}{1980}).
\newblock


\bibitem[Jha et~al\mbox{.}(2023a)]%
        {jha2023seegull}
\bibfield{author}{\bibinfo{person}{Akshita Jha}, \bibinfo{person}{Aida Davani},
  \bibinfo{person}{Chandan~K. Reddy}, \bibinfo{person}{Shachi Dave},
  \bibinfo{person}{Vinodkumar Prabhakaran}, {and} \bibinfo{person}{Sunipa
  Dev}.} \bibinfo{year}{2023}\natexlab{a}.
\newblock \showarticletitle{{SeeGULL}: A Stereotype Benchmark with Broad
  Geo-Cultural Coverage Leveraging Generative Models}.
\newblock \bibinfo{journal}{\emph{arXiv:2305.11840}} (\bibinfo{year}{2023}).
\newblock


\bibitem[Jha et~al\mbox{.}(2023b)]%
        {jha-etal-2023-seegull}
\bibfield{author}{\bibinfo{person}{Akshita Jha}, \bibinfo{person}{Aida
  Mostafazadeh~Davani}, \bibinfo{person}{Chandan~K Reddy},
  \bibinfo{person}{Shachi Dave}, \bibinfo{person}{Vinodkumar Prabhakaran},
  {and} \bibinfo{person}{Sunipa Dev}.} \bibinfo{year}{2023}\natexlab{b}.
\newblock \showarticletitle{{S}ee{GULL}: A Stereotype Benchmark with Broad
  Geo-Cultural Coverage Leveraging Generative Models}. In
  \bibinfo{booktitle}{\emph{Proceedings of the 61st Annual Meeting of the
  Association for Computational Linguistics (Volume 1: Long Papers)}},
  \bibfield{editor}{\bibinfo{person}{Anna Rogers}, \bibinfo{person}{Jordan
  Boyd-Graber}, {and} \bibinfo{person}{Naoaki Okazaki}} (Eds.).
  \bibinfo{publisher}{Association for Computational Linguistics},
  \bibinfo{address}{Toronto, Canada}, \bibinfo{pages}{9851--9870}.
\newblock
\urldef\tempurl%
\url{https://doi.org/10.18653/v1/2023.acl-long.548}
\showDOI{\tempurl}


\bibitem[Kairam and Heer(2016)]%
        {kairam2016crowds}
\bibfield{author}{\bibinfo{person}{Sanjay Kairam} {and}
  \bibinfo{person}{Jeffrey Heer}.} \bibinfo{year}{2016}\natexlab{}.
\newblock \showarticletitle{Parting Crowds: Characterizing Divergent
  Interpretations in Crowdsourced Annotation Tasks}.
\newblock \bibinfo{journal}{\emph{ACM Conference On Computer-Supported
  Cooperative Work And Social Computing (CSCW)}} (\bibinfo{year}{2016}).
\newblock


\bibitem[Kaneko and Bollegala(2019)]%
        {kaneko2019debiasing}
\bibfield{author}{\bibinfo{person}{Masahiro Kaneko} {and}
  \bibinfo{person}{Danushka Bollegala}.} \bibinfo{year}{2019}\natexlab{}.
\newblock \showarticletitle{Gender-preserving Debiasing for Pre-trained Word
  Embeddings}.
\newblock \bibinfo{journal}{\emph{Annual Conference of the Association for
  Computational Linguistics (ACL)}} (\bibinfo{year}{2019}).
\newblock


\bibitem[Karve et~al\mbox{.}(2019)]%
        {karve2019conceptor}
\bibfield{author}{\bibinfo{person}{Saket Karve}, \bibinfo{person}{Lyle Ungar},
  {and} \bibinfo{person}{João Sedoc}.} \bibinfo{year}{2019}\natexlab{}.
\newblock \showarticletitle{Conceptor Debiasing of Word Representations
  Evaluated on WEAT}.
\newblock \bibinfo{journal}{\emph{arXiv:1906.05993}} (\bibinfo{year}{2019}).
\newblock


\bibitem[Katz and Braly(1933)]%
        {katz1933racial}
\bibfield{author}{\bibinfo{person}{Daniel Katz} {and} \bibinfo{person}{Kenneth
  Braly}.} \bibinfo{year}{1933}\natexlab{}.
\newblock \showarticletitle{Racial Stereotypes of One Hundred College
  Students}.
\newblock \bibinfo{journal}{\emph{The Journal of Abnormal and Social
  Psychology}}  \bibinfo{volume}{28} (\bibinfo{year}{1933}).
\newblock
Issue 3.


\bibitem[Katzman et~al\mbox{.}(2023)]%
        {katzman2023taxonomy}
\bibfield{author}{\bibinfo{person}{Jared Katzman}, \bibinfo{person}{Angelina
  Wang}, \bibinfo{person}{Morgan Scheuerman}, \bibinfo{person}{Su~Lin
  Blodgett}, \bibinfo{person}{Kristen Laird}, \bibinfo{person}{Hanna Wallach},
  {and} \bibinfo{person}{Solon Barocas}.} \bibinfo{year}{2023}\natexlab{}.
\newblock \showarticletitle{Taxonomizing and Measuring Representational Harms:
  A Look at Image Tagging}.
\newblock \bibinfo{journal}{\emph{AAAI Conference on Artificial Intelligence}}
  (\bibinfo{year}{2023}).
\newblock


\bibitem[Kay et~al\mbox{.}(2015)]%
        {kay2015search}
\bibfield{author}{\bibinfo{person}{Matthew Kay}, \bibinfo{person}{Cynthia
  Matuszek}, {and} \bibinfo{person}{Sean~A. Munson}.}
  \bibinfo{year}{2015}\natexlab{}.
\newblock \showarticletitle{Unequal Representation and Gender Stereotypes in
  Image Search Results for Occupations}.
\newblock \bibinfo{journal}{\emph{Conference on Human Factors in Computing
  Systems (CHI)}} (\bibinfo{year}{2015}).
\newblock


\bibitem[Kukar and Kononenko(1998)]%
        {kukar1998cost}
\bibfield{author}{\bibinfo{person}{Matjaž Kukar} {and} \bibinfo{person}{Igor
  Kononenko}.} \bibinfo{year}{1998}\natexlab{}.
\newblock \showarticletitle{Cost-Sensitive Learning with Neural Networks}.
\newblock \bibinfo{journal}{\emph{European Conference on Artificial
  Intelligence}} (\bibinfo{year}{1998}).
\newblock


\bibitem[Kuznetsova et~al\mbox{.}(2020)]%
        {kuznetsova2020openimages}
\bibfield{author}{\bibinfo{person}{Alina Kuznetsova}, \bibinfo{person}{Hassan
  Rom}, \bibinfo{person}{Neil Alldrin}, \bibinfo{person}{Jasper Uijlings},
  \bibinfo{person}{Ivan Krasin}, \bibinfo{person}{Shahab~Kamali Jordi
  Pont-Tuset}, \bibinfo{person}{Stefan Popov}, \bibinfo{person}{Matteo
  Malloci}, \bibinfo{person}{Alexander Kolesnikov}, \bibinfo{person}{Tom
  Duerig}, {and} \bibinfo{person}{Vittorio Ferrari}.}
  \bibinfo{year}{2020}\natexlab{}.
\newblock \showarticletitle{The Open Images Dataset V4: Unified image
  classification, object detection, and visual relationship detection at
  scale}.
\newblock \bibinfo{journal}{\emph{International Journal of Computer Vision
  (IJCV)}} (\bibinfo{year}{2020}).
\newblock


\bibitem[Lin et~al\mbox{.}(2014)]%
        {lin2014coco}
\bibfield{author}{\bibinfo{person}{Tsung-Yi Lin}, \bibinfo{person}{Michael
  Maire}, \bibinfo{person}{Serge Belongie}, \bibinfo{person}{James Hays},
  \bibinfo{person}{Pietro Perona}, \bibinfo{person}{Deva Ramanan},
  \bibinfo{person}{Piotr Doll{\'a}r}, {and} \bibinfo{person}{C~Lawrence
  Zitnick}.} \bibinfo{year}{2014}\natexlab{}.
\newblock \showarticletitle{Microsoft {COCO}: Common objects in context}.
\newblock \bibinfo{journal}{\emph{European Conference on Computer Vision
  (ECCV)}} (\bibinfo{year}{2014}).
\newblock


\bibitem[Lippmann(1922)]%
        {lippmann1922public}
\bibfield{author}{\bibinfo{person}{Walter Lippmann}.}
  \bibinfo{year}{1922}\natexlab{}.
\newblock \showarticletitle{Public opinion.}
\newblock \bibinfo{journal}{\emph{MacMillan Co}} (\bibinfo{year}{1922}).
\newblock


\bibitem[López-Sáez and Lisbona(2014)]%
        {lopezsaez2014descriptiveprescriptive}
\bibfield{author}{\bibinfo{person}{Mercedes López-Sáez} {and}
  \bibinfo{person}{Ana Lisbona}.} \bibinfo{year}{2014}\natexlab{}.
\newblock \showarticletitle{Descriptive and prescriptive features of gender
  stereotyping. Relationships among its components}.
\newblock \bibinfo{journal}{\emph{International Journal of Social Psychology}}
  \bibinfo{volume}{24} (\bibinfo{year}{2014}).
\newblock
Issue 3.


\bibitem[Makwana et~al\mbox{.}(2018)]%
        {makwana2018transphobia}
\bibfield{author}{\bibinfo{person}{Arti~P. Makwana}, \bibinfo{person}{Kristof
  Dhont}, \bibinfo{person}{Jonas~De keersmaecker}, \bibinfo{person}{Parisa
  Akhlaghi-Ghaffarokh}, \bibinfo{person}{Marine Masure}, {and}
  \bibinfo{person}{Arne Roets}.} \bibinfo{year}{2018}\natexlab{}.
\newblock \showarticletitle{The Motivated Cognitive Basis of Transphobia: The
  Roles of Right-Wing Ideologies and Gender Role Beliefs}.
\newblock \bibinfo{journal}{\emph{Sex Roles}}  \bibinfo{volume}{79}
  (\bibinfo{year}{2018}).
\newblock


\bibitem[Manzini et~al\mbox{.}(2019)]%
        {manzini2019multiclass}
\bibfield{author}{\bibinfo{person}{Thomas Manzini}, \bibinfo{person}{Yao~Chong
  Lim}, \bibinfo{person}{Yulia Tsvetkov}, {and} \bibinfo{person}{Alan~W.
  Black}.} \bibinfo{year}{2019}\natexlab{}.
\newblock \showarticletitle{Black is to Criminal as Caucasian is to Police:
  Detecting and Removing Multiclass Bias in Word Embeddings}.
\newblock \bibinfo{journal}{\emph{Annual Conference of the North American
  Chapter of the Association for Computational Linguistics (NAACL)}}
  (\bibinfo{year}{2019}).
\newblock


\bibitem[Metaxa et~al\mbox{.}(2021)]%
        {metaxa2021image}
\bibfield{author}{\bibinfo{person}{Danaë Metaxa}, \bibinfo{person}{Michelle~A.
  Gan}, \bibinfo{person}{Su Goh}, \bibinfo{person}{Jeff Hancock}, {and}
  \bibinfo{person}{James~A. Landay}.} \bibinfo{year}{2021}\natexlab{}.
\newblock \showarticletitle{An Image of Society: Gender and Racial
  Representation and Impact in Image Search Results for Occupations}.
\newblock \bibinfo{journal}{\emph{ACM Conference on Human-Computer Interaction
  (CSCW)}} (\bibinfo{year}{2021}).
\newblock


\bibitem[Morgenroth and Ryan(2018)]%
        {morgenroth2018gendertroublesocialpsych}
\bibfield{author}{\bibinfo{person}{Thekla Morgenroth} {and}
  \bibinfo{person}{Michelle~K. Ryan}.} \bibinfo{year}{2018}\natexlab{}.
\newblock \showarticletitle{Gender Trouble in Social Psychology: How Can
  Butler’s Work Inform Experimental Social Psychologists’ Conceptualization
  of Gender?}
\newblock \bibinfo{journal}{\emph{Frontiers in Psychology}}
  (\bibinfo{year}{2018}).
\newblock


\bibitem[Morgenroth and Ryan(2020)]%
        {morgenroth2020gendertrouble}
\bibfield{author}{\bibinfo{person}{Thekla Morgenroth} {and}
  \bibinfo{person}{Michelle~K. Ryan}.} \bibinfo{year}{2020}\natexlab{}.
\newblock \showarticletitle{The Effects of Gender Trouble: An Integrative
  Theoretical Framework of the Perpetuation and Disruption of the Gender/Sex
  Binary}.
\newblock \bibinfo{journal}{\emph{Perspectives on Psychological Science}}
  \bibinfo{volume}{16} (\bibinfo{year}{2020}).
\newblock


\bibitem[Nagoshi et~al\mbox{.}(2019)]%
        {nagoshi2019prejudices}
\bibfield{author}{\bibinfo{person}{Craig~T. Nagoshi}, \bibinfo{person}{J.~Raven
  Cloud}, \bibinfo{person}{Louis~M. Lindley}, \bibinfo{person}{Julie~L.
  Nagoshi}, {and} \bibinfo{person}{Lucas~J. Lothamer }.}
  \bibinfo{year}{2019}\natexlab{}.
\newblock \showarticletitle{A Test of the Three-Component Model of Gender-Based
  Prejudices: Homophobia and Transphobia Are Affected by Raters’ and
  Targets’ Assigned Sex at Birth}.
\newblock \bibinfo{journal}{\emph{Sex Roles}}  \bibinfo{volume}{80}
  (\bibinfo{year}{2019}).
\newblock


\bibitem[Nangia et~al\mbox{.}(2020)]%
        {nangia-etal-2020-crows}
\bibfield{author}{\bibinfo{person}{Nikita Nangia}, \bibinfo{person}{Clara
  Vania}, \bibinfo{person}{Rasika Bhalerao}, {and} \bibinfo{person}{Samuel~R.
  Bowman}.} \bibinfo{year}{2020}\natexlab{}.
\newblock \showarticletitle{{C}row{S}-Pairs: A Challenge Dataset for Measuring
  Social Biases in Masked Language Models}. In
  \bibinfo{booktitle}{\emph{Proceedings of the 2020 Conference on Empirical
  Methods in Natural Language Processing (EMNLP)}},
  \bibfield{editor}{\bibinfo{person}{Bonnie Webber}, \bibinfo{person}{Trevor
  Cohn}, \bibinfo{person}{Yulan He}, {and} \bibinfo{person}{Yang Liu}} (Eds.).
  \bibinfo{publisher}{Association for Computational Linguistics},
  \bibinfo{address}{Online}, \bibinfo{pages}{1953--1967}.
\newblock
\urldef\tempurl%
\url{https://doi.org/10.18653/v1/2020.emnlp-main.154}
\showDOI{\tempurl}


\bibitem[Nicolas et~al\mbox{.}(2022)]%
        {nicolas2022spontaneous}
\bibfield{author}{\bibinfo{person}{Gandalf Nicolas}, \bibinfo{person}{Xuechunzi
  Bai}, {and} \bibinfo{person}{Susan~T Fiske}.}
  \bibinfo{year}{2022}\natexlab{}.
\newblock \showarticletitle{A spontaneous stereotype content model: Taxonomy,
  properties, and prediction.}
\newblock \bibinfo{journal}{\emph{Journal of personality and social
  psychology}} (\bibinfo{year}{2022}).
\newblock


\bibitem[Noble(2012)]%
        {noble2012crowdsourcing}
\bibfield{author}{\bibinfo{person}{Jennifer~A. Noble}.}
  \bibinfo{year}{2012}\natexlab{}.
\newblock \showarticletitle{Minority voices of crowdsourcing: why we should pay
  attention to every member of the crowd}.
\newblock \bibinfo{journal}{\emph{ACM Conference On Computer-Supported
  Cooperative Work And Social Computing (CSCW)}} (\bibinfo{year}{2012}).
\newblock


\bibitem[Noble(2018)]%
        {noble2018oppression}
\bibfield{author}{\bibinfo{person}{Safiya~Umoja Noble}.}
  \bibinfo{year}{2018}\natexlab{}.
\newblock \showarticletitle{Algorithms of Oppression: How Search Engines
  Reinforce Racism}.
\newblock \bibinfo{journal}{\emph{NYU Press}} (\bibinfo{year}{2018}).
\newblock


\bibitem[O'Dowd(2018)]%
        {odowd2018microaggression}
\bibfield{author}{\bibinfo{person}{Omaith O'Dowd}.}
  \bibinfo{year}{2018}\natexlab{}.
\newblock \showarticletitle{Microaggressions: A Kantian Account}.
\newblock \bibinfo{journal}{\emph{Ethical Theory and Moral Practice}}
  \bibinfo{volume}{21} (\bibinfo{year}{2018}).
\newblock


\bibitem[Otterbacher et~al\mbox{.}(2017)]%
        {otterbacher2017competent}
\bibfield{author}{\bibinfo{person}{Jahna Otterbacher}, \bibinfo{person}{Jo
  Bates}, {and} \bibinfo{person}{Paul Clough}.}
  \bibinfo{year}{2017}\natexlab{}.
\newblock \showarticletitle{Competent Men and Warm Women: Gender Stereotypes
  and Backlash in Image Search Results}.
\newblock \bibinfo{journal}{\emph{Conference on Human Factors in Computing
  Systems (CHI)}} (\bibinfo{year}{2017}).
\newblock


\bibitem[Otterbacher et~al\mbox{.}(2018)]%
        {otterbacher2018sexism}
\bibfield{author}{\bibinfo{person}{Jahna Otterbacher},
  \bibinfo{person}{Alessandro Checco}, \bibinfo{person}{Gianluca Demartini},
  {and} \bibinfo{person}{Paul Clough}.} \bibinfo{year}{2018}\natexlab{}.
\newblock \showarticletitle{Investigating User Perception of Gender Bias in
  Image Search: The Role of Sexism}.
\newblock \bibinfo{journal}{\emph{ACM SIGIR Conference on Research and
  Development in Information Retrieval (SIGIR)}} (\bibinfo{year}{2018}).
\newblock


\bibitem[Parrish et~al\mbox{.}(2022)]%
        {parrish-etal-2022-bbq}
\bibfield{author}{\bibinfo{person}{Alicia Parrish}, \bibinfo{person}{Angelica
  Chen}, \bibinfo{person}{Nikita Nangia}, \bibinfo{person}{Vishakh Padmakumar},
  \bibinfo{person}{Jason Phang}, \bibinfo{person}{Jana Thompson},
  \bibinfo{person}{Phu~Mon Htut}, {and} \bibinfo{person}{Samuel Bowman}.}
  \bibinfo{year}{2022}\natexlab{}.
\newblock \showarticletitle{{BBQ}: A hand-built bias benchmark for question
  answering}. In \bibinfo{booktitle}{\emph{Findings of the Association for
  Computational Linguistics: ACL 2022}},
  \bibfield{editor}{\bibinfo{person}{Smaranda Muresan},
  \bibinfo{person}{Preslav Nakov}, {and} \bibinfo{person}{Aline Villavicencio}}
  (Eds.). \bibinfo{publisher}{Association for Computational Linguistics},
  \bibinfo{address}{Dublin, Ireland}, \bibinfo{pages}{2086--2105}.
\newblock
\urldef\tempurl%
\url{https://doi.org/10.18653/v1/2022.findings-acl.165}
\showDOI{\tempurl}


\bibitem[Pałka(2023)]%
        {palka2023harm}
\bibfield{author}{\bibinfo{person}{Przemysław Pałka}.}
  \bibinfo{year}{2023}\natexlab{}.
\newblock \showarticletitle{AI, Consumers \& Psychological Harm}.
\newblock \bibinfo{journal}{\emph{Cambridge University Press}}
  (\bibinfo{year}{2023}).
\newblock


\bibitem[Ravfogel et~al\mbox{.}(2020)]%
        {ravfogel2020null}
\bibfield{author}{\bibinfo{person}{Shauli Ravfogel}, \bibinfo{person}{Yanai
  Elazar}, \bibinfo{person}{Hila Gonen}, \bibinfo{person}{Michael Twiton},
  {and} \bibinfo{person}{Yoav Goldberg}.} \bibinfo{year}{2020}\natexlab{}.
\newblock \showarticletitle{Null It Out: Guarding Protected Attributes by
  Iterative Nullspace Projection}.
\newblock \bibinfo{journal}{\emph{Annual Conference of the Association for
  Computational Linguistics (ACL)}} (\bibinfo{year}{2020}).
\newblock


\bibitem[Rini(2020)]%
        {rini2020microaggression}
\bibfield{author}{\bibinfo{person}{Regina Rini}.}
  \bibinfo{year}{2020}\natexlab{}.
\newblock \showarticletitle{The Ethics of Microaggression}.
\newblock \bibinfo{journal}{\emph{Routledge Taylr \& Francis Group}}
  (\bibinfo{year}{2020}).
\newblock


\bibitem[Rollero et~al\mbox{.}(2014)]%
        {rollero2014shortened}
\bibfield{author}{\bibinfo{person}{Chiara Rollero}, \bibinfo{person}{Peter
  Glick}, {and} \bibinfo{person}{Stefano Tartaglia}.}
  \bibinfo{year}{2014}\natexlab{}.
\newblock \showarticletitle{Psychometric properties of short versions of the
  Ambivalent Sexism Inventory and Ambivalence Toward Men Inventory}.
\newblock \bibinfo{journal}{\emph{TPM-Testing, Psychometrics, Methodology in
  Applied Psychology}}  \bibinfo{volume}{21} (\bibinfo{year}{2014}).
\newblock
Issue 2.


\bibitem[Rudman et~al\mbox{.}(2012a)]%
        {rudman2012vanguards}
\bibfield{author}{\bibinfo{person}{Laurie~A. Rudman},
  \bibinfo{person}{Corinne~A. Moss-Racusin}, \bibinfo{person}{Peter Glick},
  {and} \bibinfo{person}{Julie~E. Phelan}.} \bibinfo{year}{2012}\natexlab{a}.
\newblock \showarticletitle{Reactions to Vanguards: Advances in Backlash
  Theory}.
\newblock \bibinfo{journal}{\emph{Advances in experimental social psychology}}
  \bibinfo{volume}{45} (\bibinfo{year}{2012}).
\newblock


\bibitem[Rudman et~al\mbox{.}(2012b)]%
        {rudman2012backlash}
\bibfield{author}{\bibinfo{person}{Laurie~A. Rudman},
  \bibinfo{person}{Corinne~A. Moss-Racusin}, \bibinfo{person}{Julie~E. Phelan},
  {and} \bibinfo{person}{Sanne Nauts}.} \bibinfo{year}{2012}\natexlab{b}.
\newblock \showarticletitle{Status incongruity and backlash effects: Defending
  the gender hierarchy motivates prejudice against female leaders}.
\newblock \bibinfo{journal}{\emph{Journal of Experimental Social Psychology}}
  \bibinfo{volume}{48} (\bibinfo{year}{2012}).
\newblock


\bibitem[Savoldi et~al\mbox{.}(2021)]%
        {savoldi-etal-2021-gender}
\bibfield{author}{\bibinfo{person}{Beatrice Savoldi}, \bibinfo{person}{Marco
  Gaido}, \bibinfo{person}{Luisa Bentivogli}, \bibinfo{person}{Matteo Negri},
  {and} \bibinfo{person}{Marco Turchi}.} \bibinfo{year}{2021}\natexlab{}.
\newblock \showarticletitle{Gender Bias in Machine Translation}.
\newblock \bibinfo{journal}{\emph{Transactions of the Association for
  Computational Linguistics}}  \bibinfo{volume}{9} (\bibinfo{year}{2021}),
  \bibinfo{pages}{845--874}.
\newblock
\urldef\tempurl%
\url{https://doi.org/10.1162/tacl_a_00401}
\showDOI{\tempurl}


\bibitem[Schumann et~al\mbox{.}(2021)]%
        {schumann2021miap}
\bibfield{author}{\bibinfo{person}{Candice Schumann}, \bibinfo{person}{Susanna
  Ricco}, \bibinfo{person}{Utsav Prabhu}, \bibinfo{person}{Vittorio Ferrari},
  {and} \bibinfo{person}{Caroline Pantofaru}.} \bibinfo{year}{2021}\natexlab{}.
\newblock \showarticletitle{A Step Toward More Inclusive People Annotations for
  Fairness}.
\newblock \bibinfo{journal}{\emph{ACM Conference on Artificial Intelligence,
  Ethics, and Society (AIES)}} (\bibinfo{year}{2021}).
\newblock


\bibitem[Seger et~al\mbox{.}(2017)]%
        {seger2017emotion}
\bibfield{author}{\bibinfo{person}{Charles~R. Seger}, \bibinfo{person}{Ishani
  Banerji}, \bibinfo{person}{Sang~Hee Park}, \bibinfo{person}{Eliot~R. Smith},
  {and} \bibinfo{person}{Diane~M. Mackie}.} \bibinfo{year}{2017}\natexlab{}.
\newblock \showarticletitle{Specific emotions as mediators of the effect of
  intergroup contact on prejudice: findings across multiple participant and
  target groups}.
\newblock \bibinfo{journal}{\emph{Cognition and Emotion}}  \bibinfo{volume}{31}
  (\bibinfo{year}{2017}).
\newblock
Issue 5.


\bibitem[Selvaraju et~al\mbox{.}(2017)]%
        {selvaraju2017grad}
\bibfield{author}{\bibinfo{person}{Ramprasaath~R Selvaraju},
  \bibinfo{person}{Michael Cogswell}, \bibinfo{person}{Abhishek Das},
  \bibinfo{person}{Ramakrishna Vedantam}, \bibinfo{person}{Devi Parikh}, {and}
  \bibinfo{person}{Dhruv Batra}.} \bibinfo{year}{2017}\natexlab{}.
\newblock \showarticletitle{Grad-cam: Visual explanations from deep networks
  via gradient-based localization}. In \bibinfo{booktitle}{\emph{Proceedings of
  the IEEE international conference on computer vision}}.
  \bibinfo{pages}{618--626}.
\newblock


\bibitem[Shin et~al\mbox{.}(2020)]%
        {shin2020neutralizing}
\bibfield{author}{\bibinfo{person}{Seungjae Shin}, \bibinfo{person}{Kyungwoo
  Song}, \bibinfo{person}{JoonHo Jang}, \bibinfo{person}{Hyemi Kim},
  \bibinfo{person}{Weonyoung Joo}, {and} \bibinfo{person}{Il-Chul Moon}.}
  \bibinfo{year}{2020}\natexlab{}.
\newblock \showarticletitle{Neutralizing Gender Bias in Word Embedding with
  Latent Disentanglement and Counterfactual Generation}.
\newblock \bibinfo{journal}{\emph{Findings of EMNLP}} (\bibinfo{year}{2020}).
\newblock


\bibitem[Shnabel et~al\mbox{.}(2016)]%
        {shnabel2016help}
\bibfield{author}{\bibinfo{person}{Nurit Shnabel}, \bibinfo{person}{Yoav
  Bar-Anan}, \bibinfo{person}{Anna Kende}, \bibinfo{person}{Orly Bareket},
  {and} \bibinfo{person}{Yael Lazar}.} \bibinfo{year}{2016}\natexlab{}.
\newblock \showarticletitle{Help to perpetuate traditional gender roles:
  Benevolent sexism increases engagement in dependency-oriented cross-gender
  helping.}
\newblock \bibinfo{journal}{\emph{Journal of personality and social
  psychology}} \bibinfo{volume}{110}, \bibinfo{number}{1}
  (\bibinfo{year}{2016}), \bibinfo{pages}{55}.
\newblock


\bibitem[Simonite(2018)]%
        {simonite2018comes}
\bibfield{author}{\bibinfo{person}{Tom Simonite}.}
  \bibinfo{year}{2018}\natexlab{}.
\newblock \showarticletitle{{When It Comes to Gorillas, Google Photos Remains
  Blind}}.
\newblock \bibinfo{journal}{\emph{Wired, January}} (\bibinfo{year}{2018}).
\newblock


\bibitem[Sotnikova et~al\mbox{.}(2021)]%
        {sotnikova2021stereotype}
\bibfield{author}{\bibinfo{person}{Anna Sotnikova},
  \bibinfo{person}{Yang~Trista Cao}, \bibinfo{person}{Hal~Daumé III}, {and}
  \bibinfo{person}{Rachel Rudinger}.} \bibinfo{year}{2021}\natexlab{}.
\newblock \showarticletitle{Analyzing Stereotypes in Generative Text Inference
  Tasks}.
\newblock \bibinfo{journal}{\emph{Findings of the Association for Computational
  Linguistics: ACL-IJCNLP}} (\bibinfo{year}{2021}).
\newblock


\bibitem[Spencer et~al\mbox{.}(2015)]%
        {spencer2015threat}
\bibfield{author}{\bibinfo{person}{Steven~J. Spencer},
  \bibinfo{person}{Christine Logel}, {and} \bibinfo{person}{Paul~G. Davies}.}
  \bibinfo{year}{2015}\natexlab{}.
\newblock \showarticletitle{Stereotype Threat}.
\newblock \bibinfo{journal}{\emph{Annual Review of Psychology}}
  \bibinfo{volume}{67} (\bibinfo{year}{2015}).
\newblock


\bibitem[Srivastava et~al\mbox{.}(2019)]%
        {srivastava2019perception}
\bibfield{author}{\bibinfo{person}{Megha Srivastava}, \bibinfo{person}{Hoda
  Heidari}, {and} \bibinfo{person}{Andreas Krause}.}
  \bibinfo{year}{2019}\natexlab{}.
\newblock \showarticletitle{Mathematical Notions vs. Human Perception of
  Fairness: A Descriptive Approach to Fairness for Machine Learning}.
\newblock \bibinfo{journal}{\emph{ACM SIGKDD International Conference on
  Knowledge Discovery and Data Mining (KDD)}} (\bibinfo{year}{2019}).
\newblock


\bibitem[Steele and Aronson(1995)]%
        {steele1996stereotypethreat}
\bibfield{author}{\bibinfo{person}{Claude~M. Steele} {and}
  \bibinfo{person}{Joshua Aronson}.} \bibinfo{year}{1995}\natexlab{}.
\newblock \showarticletitle{Stereotype threat and the intellectual test
  performance of African Americans}.
\newblock \bibinfo{journal}{\emph{Journal of Personality and Social
  Psychology}}  \bibinfo{volume}{69} (\bibinfo{year}{1995}).
\newblock
Issue 5.


\bibitem[Stern and Rule(2017)]%
        {stern2017androgyny}
\bibfield{author}{\bibinfo{person}{Chadly Stern} {and}
  \bibinfo{person}{Nicholas~O. Rule}.} \bibinfo{year}{2017}\natexlab{}.
\newblock \showarticletitle{Physical Androgyny and Categorization Difficulty
  Shape Political Conservatives’ Attitudes Toward Transgender People}.
\newblock \bibinfo{journal}{\emph{Social Psychological and Personality
  Science}} (\bibinfo{year}{2017}).
\newblock


\bibitem[van Miltenburg(2016)]%
        {miltenburg2016stereotype}
\bibfield{author}{\bibinfo{person}{Emiel van Miltenburg}.}
  \bibinfo{year}{2016}\natexlab{}.
\newblock \showarticletitle{Stereotyping and Bias in the Flickr30K Dataset}.
\newblock \bibinfo{journal}{\emph{Proceedings of the Workshop on Multimodal
  Corpora}} (\bibinfo{year}{2016}).
\newblock


\bibitem[Vandello et~al\mbox{.}(2008)]%
        {vandello2008precarious}
\bibfield{author}{\bibinfo{person}{Joseph~A. Vandello},
  \bibinfo{person}{Jennifer~K. Bosson}, \bibinfo{person}{Dov Cohen},
  \bibinfo{person}{Burnaford Rochelle~M}, {and} \bibinfo{person}{Jonathan~R.
  Weaver}.} \bibinfo{year}{2008}\natexlab{}.
\newblock \showarticletitle{Precarious manhood}.
\newblock \bibinfo{journal}{\emph{Journal of Personality and Social
  Psychology}} (\bibinfo{year}{2008}).
\newblock


\bibitem[Vlasceanu and Amodio(2022)]%
        {vlasceanu2022internet}
\bibfield{author}{\bibinfo{person}{Madalinea Vlasceanu} {and}
  \bibinfo{person}{David~M. Amodio}.} \bibinfo{year}{2022}\natexlab{}.
\newblock \showarticletitle{Propagation of societal gender inequality by
  internet search algorithms}.
\newblock \bibinfo{journal}{\emph{Proceedings of the National Academy of
  Sciences of the United States of America (PNAS)}}  \bibinfo{volume}{119}
  (\bibinfo{year}{2022}).
\newblock


\bibitem[Wan et~al\mbox{.}(2023)]%
        {wan2023kellywarm}
\bibfield{author}{\bibinfo{person}{Yixin Wan}, \bibinfo{person}{George Pu},
  \bibinfo{person}{Jiao Sun}, \bibinfo{person}{Aparna Garimella},
  \bibinfo{person}{Kai-Wei Chang}, {and} \bibinfo{person}{Nanyun Peng}.}
  \bibinfo{year}{2023}\natexlab{}.
\newblock \showarticletitle{"Kelly is a Warm Person, Joseph is a Role Model":
  Gender Biases in LLM-Generated Reference Letters}.
\newblock \bibinfo{journal}{\emph{Conference on Empirical Methods in Natural
  Language Processing (EMNLP Findings)}} (\bibinfo{year}{2023}).
\newblock


\bibitem[Wang et~al\mbox{.}(2022)]%
        {revisetool_extended}
\bibfield{author}{\bibinfo{person}{Angelina Wang}, \bibinfo{person}{Alexander
  Liu}, \bibinfo{person}{Ryan Zhang}, \bibinfo{person}{Anat Kleiman},
  \bibinfo{person}{Leslie Kim}, \bibinfo{person}{Dora Zhao},
  \bibinfo{person}{Iroha Shirai}, \bibinfo{person}{Arvind Narayanan}, {and}
  \bibinfo{person}{Olga Russakovsky}.} \bibinfo{year}{2022}\natexlab{}.
\newblock \showarticletitle{{REVISE}: A Tool for Measuring and Mitigating Bias
  in Visual Datasets}.
\newblock \bibinfo{journal}{\emph{International Journal of Computer Vision
  (IJCV)}} (\bibinfo{year}{2022}).
\newblock


\bibitem[Wang and Russakovsky(2021)]%
        {wang2021biasamp}
\bibfield{author}{\bibinfo{person}{Angelina Wang} {and} \bibinfo{person}{Olga
  Russakovsky}.} \bibinfo{year}{2021}\natexlab{}.
\newblock \showarticletitle{Directional Bias Amplification}.
\newblock \bibinfo{journal}{\emph{International Conference on Machine Learning
  (ICML)}} (\bibinfo{year}{2021}).
\newblock


\bibitem[Wang et~al\mbox{.}(2021)]%
        {wang2021acceptance}
\bibfield{author}{\bibinfo{person}{Clarice Wang}, \bibinfo{person}{Kathryn
  Wang}, \bibinfo{person}{Andrew Bian}, \bibinfo{person}{Rashidul Islam},
  \bibinfo{person}{Kamrun~Naher Keya}, \bibinfo{person}{James Foulds}, {and}
  \bibinfo{person}{Shimei Pan}.} \bibinfo{year}{2021}\natexlab{}.
\newblock \showarticletitle{User Acceptance of Gender Stereotypes in Automated
  Career Recommendations}.
\newblock \bibinfo{journal}{\emph{arXiv:2106.07112}} (\bibinfo{year}{2021}).
\newblock


\bibitem[Wang et~al\mbox{.}(2019)]%
        {wang2019balanced}
\bibfield{author}{\bibinfo{person}{Tianlu Wang}, \bibinfo{person}{Jieyu Zhao},
  \bibinfo{person}{Mark Yatskar}, \bibinfo{person}{Kai-Wei Chang}, {and}
  \bibinfo{person}{Vicente Ordonez}.} \bibinfo{year}{2019}\natexlab{}.
\newblock \showarticletitle{Balanced Datasets Are Not Enough: Estimating and
  Mitigating Gender Bias in Deep Image Representations}.
\newblock \bibinfo{journal}{\emph{International Conference on Computer Vision
  (ICCV)}} (\bibinfo{year}{2019}).
\newblock


\bibitem[Waseem(2016)]%
        {waseem2016racist}
\bibfield{author}{\bibinfo{person}{Zeerak Waseem}.}
  \bibinfo{year}{2016}\natexlab{}.
\newblock \showarticletitle{Are You a Racist or Am I Seeing Things? Annotator
  Influence on Hate Speech Detection on Twitter}.
\newblock \bibinfo{journal}{\emph{Proceedings of the First Workshop on NLP and
  Computational Social Science}} (\bibinfo{year}{2016}).
\newblock


\bibitem[Watson et~al\mbox{.}(1988)]%
        {watson1988panas}
\bibfield{author}{\bibinfo{person}{David Watson}, \bibinfo{person}{Lee~Anna
  Clark}, {and} \bibinfo{person}{Auke Tellegen}.}
  \bibinfo{year}{1988}\natexlab{}.
\newblock \showarticletitle{Development and validation of brief measures of
  positive and negative affect: the PANAS scales}.
\newblock \bibinfo{journal}{\emph{Journal of Personality and Social
  Psychology}}  \bibinfo{volume}{54} (\bibinfo{year}{1988}).
\newblock


\bibitem[Williams and Best(1977)]%
        {williams1977adjective}
\bibfield{author}{\bibinfo{person}{John~E. Williams} {and}
  \bibinfo{person}{Deborah~L. Best}.} \bibinfo{year}{1977}\natexlab{}.
\newblock \showarticletitle{Sex Stereotypes and Trait Favorability on the
  Adjective Check List}.
\newblock \bibinfo{journal}{\emph{Educational and Psychological Measurement}}
  \bibinfo{volume}{37} (\bibinfo{year}{1977}).
\newblock
Issue 1.


\bibitem[Woodruff et~al\mbox{.}(2018)]%
        {woodruff2018qualitative}
\bibfield{author}{\bibinfo{person}{Allison Woodruff}, \bibinfo{person}{Sarah~E.
  Fox}, \bibinfo{person}{Steven Rousso-Schindler}, {and}
  \bibinfo{person}{Jeffrey Warshaw}.} \bibinfo{year}{2018}\natexlab{}.
\newblock \showarticletitle{A Qualitative Exploration of Perceptions of
  Algorithmic Fairness}.
\newblock \bibinfo{journal}{\emph{Conference on Human Factors in Computing
  Systems (CHI)}} (\bibinfo{year}{2018}).
\newblock


\bibitem[Wylie(2003)]%
        {wylie2003standpoint}
\bibfield{author}{\bibinfo{person}{Alison Wylie}.}
  \bibinfo{year}{2003}\natexlab{}.
\newblock \showarticletitle{Why Standpoint Matters}.
\newblock \bibinfo{journal}{\emph{Science and Other Cultures: Issues in
  Philosophies of Science and Technology}} (\bibinfo{year}{2003}).
\newblock


\bibitem[Yaghini et~al\mbox{.}(2021)]%
        {yaghini2021human}
\bibfield{author}{\bibinfo{person}{Mohammad Yaghini}, \bibinfo{person}{Andreas
  Krause}, {and} \bibinfo{person}{Hoda Heidari}.}
  \bibinfo{year}{2021}\natexlab{}.
\newblock \showarticletitle{A Human-in-the-loop Framework to Construct
  Context-aware Mathematical Notions of Outcome Fairness}.
\newblock \bibinfo{journal}{\emph{AAAI/ACM Conference on Artificial
  Intelligence, Ethics, and Society}} (\bibinfo{year}{2021}).
\newblock


\bibitem[Zhao et~al\mbox{.}(2021)]%
        {zhao2021caption}
\bibfield{author}{\bibinfo{person}{Dora Zhao}, \bibinfo{person}{Angelina Wang},
  {and} \bibinfo{person}{Olga Russakovsky}.} \bibinfo{year}{2021}\natexlab{}.
\newblock \showarticletitle{Understanding and Evaluating Racial Biases in Image
  Captioning}.
\newblock \bibinfo{journal}{\emph{International Conference on Computer Vision
  (ICCV)}} (\bibinfo{year}{2021}).
\newblock


\bibitem[Zhao et~al\mbox{.}(2017)]%
        {zhao2017men}
\bibfield{author}{\bibinfo{person}{Jieyu Zhao}, \bibinfo{person}{Tianlu Wang},
  \bibinfo{person}{Mark Yatskar}, \bibinfo{person}{Vicente Ordonez}, {and}
  \bibinfo{person}{Kai-Wei Chang}.} \bibinfo{year}{2017}\natexlab{}.
\newblock \showarticletitle{{Men also like shopping: Reducing gender bias
  amplification using corpus-level constraints}}.
\newblock \bibinfo{journal}{\emph{Proceedings of the Conference on Empirical
  Methods in Natural Language Processing (EMNLP)}} (\bibinfo{year}{2017}).
\newblock


\bibitem[Zhao et~al\mbox{.}(2018a)]%
        {zhao2018winobias}
\bibfield{author}{\bibinfo{person}{Jieyu Zhao}, \bibinfo{person}{Tianlu Wang},
  \bibinfo{person}{Mark Yatskar}, \bibinfo{person}{Vicente Ordonez}, {and}
  \bibinfo{person}{Kai-Wei Chang}.} \bibinfo{year}{2018}\natexlab{a}.
\newblock \showarticletitle{Gender Bias in Coreference Resolution: Evaluation
  and Debiasing Methods}.
\newblock \bibinfo{journal}{\emph{North American Chapter of the Association for
  Computational Linguistics}} (\bibinfo{year}{2018}).
\newblock


\bibitem[Zhao et~al\mbox{.}(2018b)]%
        {zhao2018gnglove}
\bibfield{author}{\bibinfo{person}{Jieyu Zhao}, \bibinfo{person}{Yichao Zhou},
  \bibinfo{person}{Zeyu Li}, \bibinfo{person}{Wei Wang}, {and}
  \bibinfo{person}{Kai-Wei Chang}.} \bibinfo{year}{2018}\natexlab{b}.
\newblock \showarticletitle{Learning Gender-Neutral Word Embeddings}.
\newblock \bibinfo{journal}{\emph{Empirical Methods in Natural Language
  Processing (EMNLP)}} (\bibinfo{year}{2018}).
\newblock


\end{thebibliography}

\appendix
\section{Methods Details}
\label{app:methods}
Here we present additional details on our methods.

\subsection{Participants}
While men and women generally tend to hold the same gender stereotypes~\cite{williams1977adjective,lopezsaez2014descriptiveprescriptive, hentschel2019dimensions, eagly2020temporal}, we still collect equal numbers of participants who identify as men and women, and use this variable as a covariate throughout. Due to limitations in the survey platform which only allow us to specify gender as ``male'' or ``female,'' this formulation excludes people who identify as non-binary, which is a harmful limitation. 
Because we do not control for race in the recruitment of participants, our sample diverges from a nationally representative sample. For the gender stereotype scope of our current work, we find this to be an acceptable limitation, especially given that one defining feature of stereotypes is they are largely shared through a cultural consensus~\cite{katz1933racial}.

We did not use quality check questions in any of our surveys, because our pilot studies showed high quality responses. Instead, we used filters on Cloud Research to only recruit participants who have had at least 50 HITs approved, and have a HIT approval rate of 98\%.

\subsection{Studies 1, 4: Distinguishing Errors by Stereotype}
When asking about which machine learning errors are stereotypes, we make sure to ask participants about their perception of stereotypes held by Americans, rather than for their personal beliefs~\cite{devine1995trilogy}. The Study 1 text is ``In this survey, you will answer questions about 20 objects and their associated stereotypes. A stereotype is an overgeneralized belief about a group of people. Please note that we are not interested in your personal beliefs, but in how you think these groups are viewed by American Society. For each object, please indicate which demographic of men or women are stereotypically associated with the object. If no group is, only select the answer choice for “neither.” If both groups are, then select both groups. As an example, ping pong balls are sometimes stereotypically associated with Asian people.''

\subsection{Study 3a: Measuring Pragmatic Harm}
\begin{figure*}[h]
    \centering
    \includegraphics[width=0.9\textwidth]{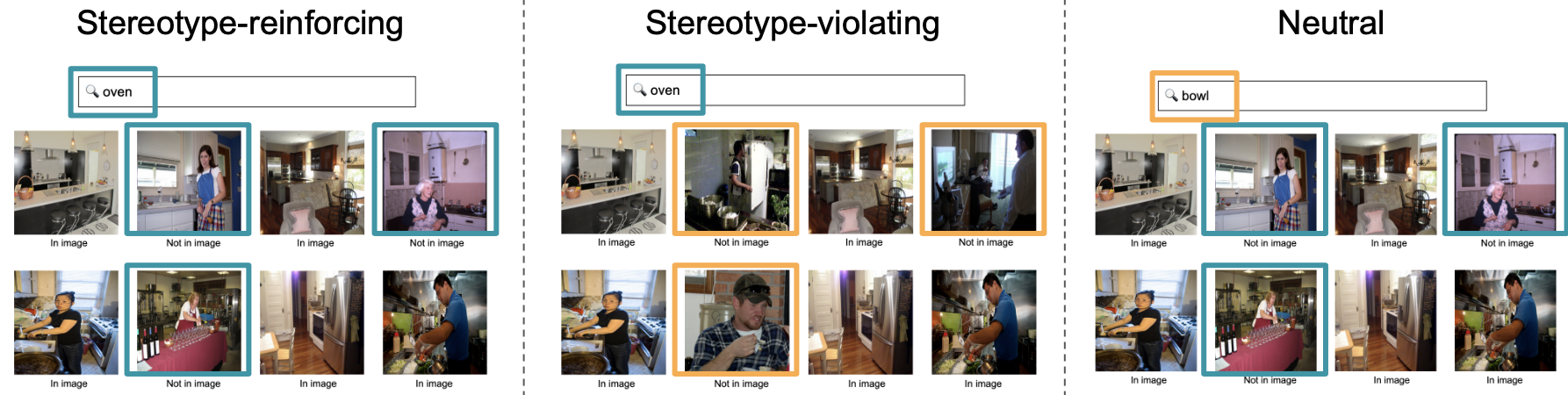}
    \caption{\textbf{Study 3 Stimuli.} Our three different stimuli are shown for the conditions: stereotype-reinforcing, stereotype-violating, and neutral. They are all image search results containing minimal changes from each other, each of which indicates whether the search query is pictured in the image, i.e., if the image search retrieval was correct or not. The teal and orange squares indicate that the only difference between the stimuli, as all images which contain an oven also contain a bowl, and all which do not contain an oven also do not contain a bowl. This was a deliberate choice to control for all potential confounding factors from the images in the study.}
    \label{fig:prag_simul}
\end{figure*}

We conduct a between-subjects survey experiment on participants who are exposed to an image search result page that contain one of three types of errors: stereotype-reinforcing, stereotype-violating, or neutral (Fig.~\ref{fig:prag_simul}).\footnote{The people pictured in our search results pages are predominantly White, which is the majority group in the dataset we employ.} To have the participants engage with these results we ask them to describe it in 3-4 sentences. Next, we ask them the behavior questions, then re-expose them to the stimulus before asking them the cognitive belief and attitude questions. We analyze changes in cognitive beliefs, attitudes, and behaviors as pragamatic harms resulting from stereotype-reinforcing errors compared to the two other conditions as controls. In this section when describing our method, we will use as examples \texttt{oven} and women for the stereotype-reinforcing error, \texttt{oven} and men for the stereotype-violating error, and \texttt{bowl} and women for the neutral one. Each question we ask is carefully grounded in the social psychology literature.


The stimuli take the form of an image search result and are pictured in Fig.~\ref{fig:prag_simul} with teal and orange colored boxes around the component of the image that changes between conditions. The search bar contains the search query, and then eight images that may or may not be correctly retrieved are shown. Each of the eight images is annotated with either ``In image'' or ``Not in image'' to make it clear to the participant which images are correct or not. The stereotype-reinforcing condition on the left contains the search query of ``oven'' with five correctly identified ovens, and three false positive images that all contain women. In other words, this classifier erroneously (and stereotypically) assumes there are ovens in images of women. The stereotype-violating condition contains the same search query, but the mistakes are replaced with false positive images that all contain men. The neutral condition contains all of the exact same images as the stereotype-reinforcing condition, with the only change being that the search query is now ``bowl'' instead of ``oven.'' This is because the five correct images were deliberately chosen to contain both bowls and ovens, which allows us to control for the variance between the different search conditions. All false positive images were selected from the actual errors of a Vision Transformer (ViT) model~\cite{dosovitskiy2021vit} trained on COCO so that they are as realistic as possible to a computer vision model's errors, and not completely egregious, e.g., a picture of a woman in a sports field as a false positive for ``oven'' or ``bowl.''

\begin{table*}
\caption{The time, pay, and reported races of the participants for each of our five studies. The full column names of races from left to right are: American Indian or Alaska Native, Asian, Black or African American, Hispanic or Latinx, Native Hawaiian or Other Pacific Islander, White, Multi-Racial / Other, and Prefer not to say.}
\label{tbl:demographics}
\begin{tabular}{ >{\centering\arraybackslash}p{1.cm}>{\centering\arraybackslash}p{.5cm}>{\centering\arraybackslash}p{.5cm}>{\centering\arraybackslash}p{.7cm}>{\centering\arraybackslash}p{.9cm}>{\centering\arraybackslash}p{.7cm}>{\centering\arraybackslash}p{.7cm}>{\centering\arraybackslash}p{.7cm}>{\centering\arraybackslash}p{.7cm}>{\centering\arraybackslash}p{.7cm}>{\centering\arraybackslash}p{.7cm}>{\centering\arraybackslash}p{.7cm}>{\centering\arraybackslash}p{.6cm} } 
 \hline
  Study  & Time (min) & Pay (\$) & Gender & AI/AN & Asian & Black & H/L & NHOPI & White & MR/O & PNTS & \textbf{Total} \\ 
 \hline
1 and 2 & 7 & 1.75 & Women & 0 & 3 & 5 & 0 & 0 & 25  & 6 & 1 & 40 \\ \cmidrule{4-13}
&& &  Men & 1 & 4 & 2 & 2 & 0 & 30 & 1 & 0 & 40\\ \hline

3a & 10 & 2.50 & Women & 1 & 11 & 32 & 8 & 0 & 229 & 19 & 0 & 300 \\ \cmidrule{4-13}
 & & & Men & 0 & 19 & 35 & 10 & 1 & 211 & 22 & 2 & 300\\ \hline

3b & 5 & 1.25 & Women & 0 & 4 & 7 & 3 & 1 & 35 & 5 & 0 & 50\\ \cmidrule{4-13}
 & & & Men & 0 & 4 & 2 & 3 & 1 & 35 & 5 & 0 & 50 \\ \hline

4  & 4 & 1 &Women & 0 & 5 & 8 & 0 & 0 & 42 & 4 & 1 & 60\\ \cmidrule{4-13}
(Labeling)&& & Men & 0 & 2 & 6 & 5 & 1 & 44 & 2 & 0& 60\\ \hline

 4 & 5 & 1.25 & Women & 0 & 5 & 15 & 1 & 0 & 120 & 7 & 2 & 150\\ \cmidrule{4-13}
(Harms) & & & Men & 1 & 9 & 17 & 6 & 1 & 107 & 9 & 0 & 150 \\ \hline

\end{tabular}
\end{table*}

\begin{figure*}[h]
    \centering
    \includegraphics[width=0.9\textwidth]{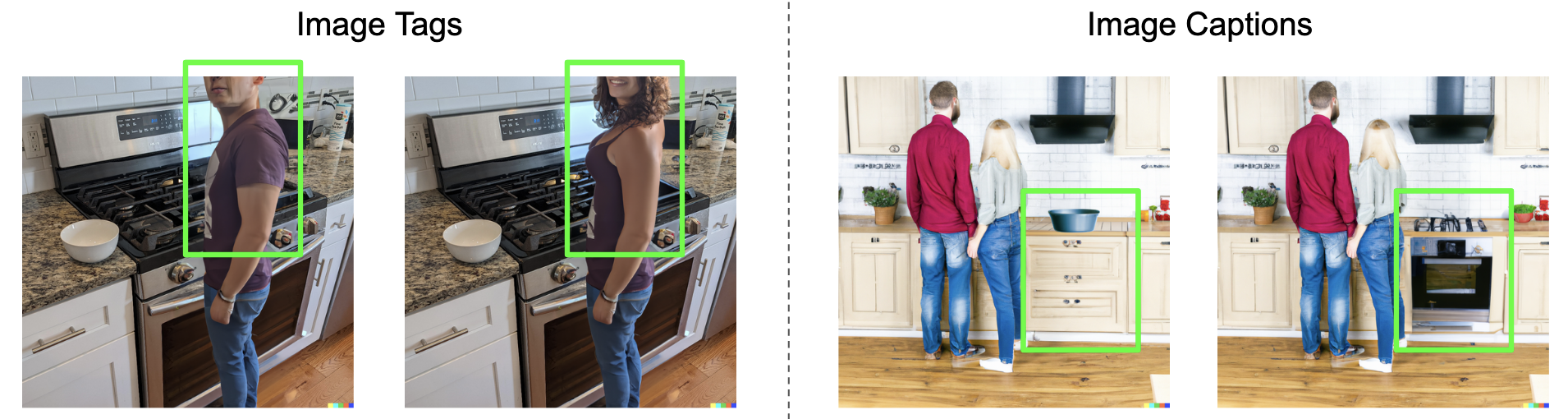}
    \caption{To measure behavioral tendencies, we ask participants to complete a realistic data annotation task on images which are created and manipulated by DALL-E2. The left pair is for the annotation of image tags, and the right pair is for image captions. Each participant is shown one image from each pair, and then we perform a between-subjects analysis to understand whether perceived gender expression affects the tags, and whether object shown influences how people of different perceived genders are described.}
    \label{fig:behavior_task}
\end{figure*}

\subsubsection{Dependent Variables}
For most of our measurements, we simply use the measure directly (e.g., the value for competence of women) as the dependent variable to regress on. For the measurements that we do something more complicated, we describe below.

\textbf{Behavior - Tags. }
Each participant produces a set of three ordered tags associated with an image of a feminine-presenting person and a set associated with a counterfactual image of a masculine-presenting person. We convert this set of tags by scoring the presence of the object in question, e.g., ``hair dryer'' (along with common misspellings such as ``hair drier'') based on its position in the ordered list of tags. When the word is present in the first spot it is given 3 points, second spot 2 points, third spot 1 point, otherwise no points.
The dependent variable is the score of both the stereotypical and neutral object on the feminine-presenting person. This is intended to capture whether the stereotype-reinforcing condition is able to increase the presence of the stereotype tag more than just the priming effect captured by the neutral object.

\textbf{Behavior - Captions.}
We offer some descriptive statistics about the captions in Appendix~\ref{app:caption_results}. This analysis was mostly exploratory, and we do not find any statistically significant differences. We first ran Study 3a looking at pragmatic harms on the stereotype of women and \texttt{oven} (with \texttt{bowl} as the control). In this iteration, we asked that respondents please describe each person in the image in separate sentences. However, there was too much noise in how respondents interpreted this set of instructions, such that the data became hard to interpret. Thus, in our second iteration of this study using the stereotype of women and hair dryer (with \texttt{toothbrush} as the control), we have two separate text entry boxes to caption each person in the image. We only present the results of this iteration in the table, as we were unable to parse anything differentiating in the first iteration.

\textbf{Cognitive - Object Use.}
In this measurement, we have a value from -10 (mostly men) to 10 (mostly women) for both the stereotypical and neutral object.
The dependent variable is the summation of both values. Again, this is intended to capture whether the stereotype-reinforcing condition is able to change the value of its associated object more than the control condition is able to.

\subsection{Study 3b, 4: Measuring Experimental Harm}
In Study 3b, in addition to personal discomfort, we also ask about societal harm. This way, even if the participant does not personally feel harmed, they may feel it on behalf of the stereotyped group. However, we find that participants' responses to both personal and societal harm are extremely correlated, and leave the results for the latter in Appendix~\ref{app:3b}.

\section{Sample Size Justification}

Our sample size selection method is recorded on Open Science Framework and is done as follows: Study 1 we selected the number such that each COCO object was labeled by 10 participants from each gender; once we saw there was sufficient consensus from Study 1, for the first part of Study 4 where we labeled OpenImages objects, we selected the number such that each OpenImages object was labeled by 5 participants from each gender; Study 3a we had three stimulus conditions across two objects, so for this between-subjects study selected the number to have 50 participants from each gender for each object-condition setting; Study 3b we had three stimulus conditions across four objects, but this is a within-subjects study so each participant sees all possible scenarios, and thus we again selected the number to have 50 participants from each gender; Study 4 we had 40 objects and as our last study ended up having the budget to have around 37.5 participants per object.

\section{Additional Results from Study 1}
\label{app:study1_results}
\begin{figure*}[h]
    \centering
    \includegraphics[width=0.9\textwidth]{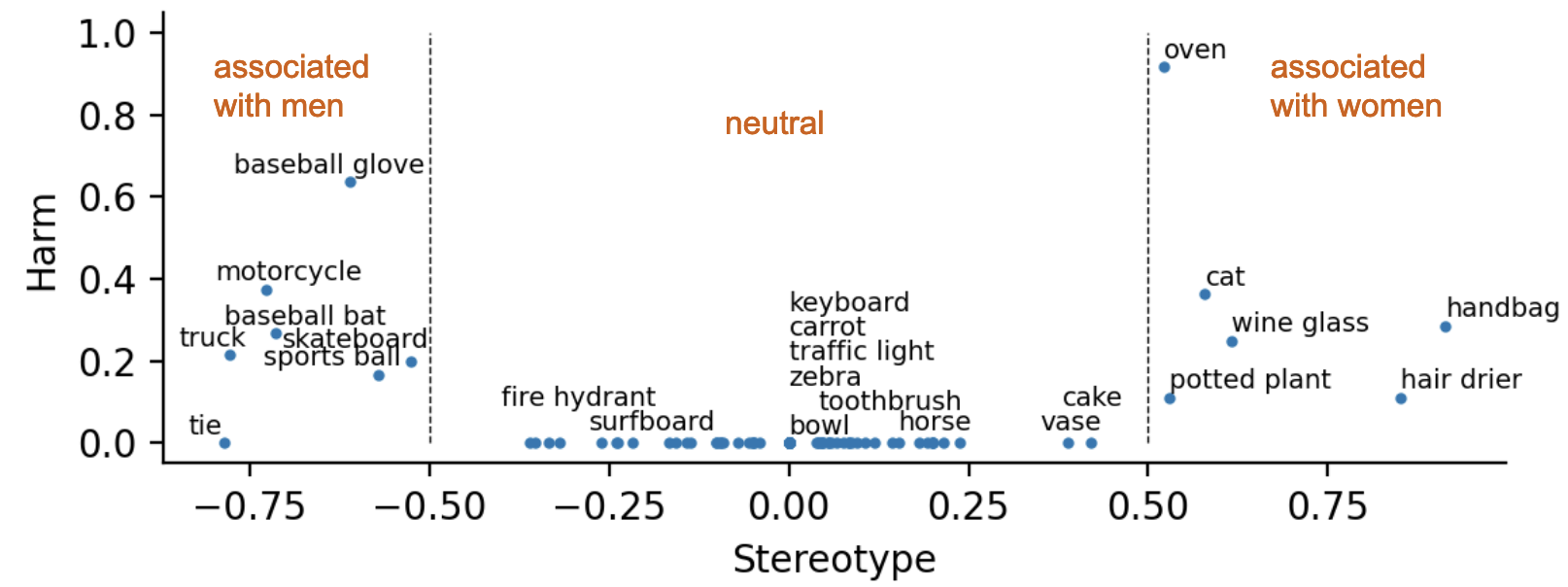}
    \caption{\textbf{Study 1 Results.} Participant responses for 80 objects in COCO dataset. The x-axis indicates the percentage of participants who indicated an object is a stereotype, where negative numbers indicate it is a stereotype about men, and positive numbers about women. For objects where more than half of the respondents indicate it is a stereotype, the y-axis indicates the percentage who marked it to be harmful.}
    \label{fig:study1_sum}
\end{figure*}

We show the results of harm annotation in the abstract on the y-axis of Fig.~\ref{fig:study1_sum} for the 13 objects marked as stereotypes. We see large variations within stereotypical objects for whether the association is perceived to be harmful when it is disconnected from a particular impact.

\section{Additional Results from Study 2}
\label{app:study2_results}

Here we present the full analyses we perform on the open-ended responses we received in Study 2 regarding why participants believe an object is a stereotype, and if so, why they find it harmful or not.

Our categorization for why an object is a stereotype or not are as follows (some responses did not fall into any of the categories):
\begin{itemize}
    \item Descriptive (45\%), e.g., for handbag and women: ``women are often seen wearing handbags and buying them''
    \item Occupation/role (22\%), e.g., for oven and women: ``Women are stereotyped to always be in the kitchen cooking while the men go out and work''
    \item Trait (11\%), e.g., for chair and men: ``sometimes men would be seen as coming home and just being lazy and lounging in their chair''
    \item Pop culture (8\%), e.g., for cow and women: ``Most people who describe a women as a cow are being harmful and hurtful''
    \item Connection to another object (5\%), e.g., for vase and women: ``I think women are seen as liking flowers, which are often put into a vase''
    \item Prescriptive (3\%), e.g., for handbag and women: ``society generally believes that only women should carry handbags''
\end{itemize}

Our categorization for why a stereotypical object is harmful is as follows (some responses did not fall into any of the categories):
\begin{itemize}
    \item Proscriptive, i.e., excluding (40\%), e.g., for dining table and women: ``it makes it looked down upon if a man cooks dinner.''
    \item Prescriptive, i.e., restricting (26\%), e.g., for dining table and women: ``I think it puts women in a box that says they must prepare dinner''
    \item Negative Trait (13\%), e.g., for handbag and women: ``It is harmful because it implies that women cares more about looks and their appearance.''
    \item Demeans (10\%), e.g., for cow and women: ``cow is a typical insult for a women a man doesn't like  (`stupid cow')''
    \item Objectifies (5\%), e.g., for cup and women: ``It is harmful because a cup is an object and it's comparing it to a woman''
    \item Sexism (3\%), e.g., for sandwich and women: ``make me a sandwich meme sexism''
    \item Incorrect (3\%), e.g., for sandwich and women: ``It's an old, tired stereotype that holds no merit.''
\end{itemize}

The following are the two reasons respondents listed a stereotype to not be harmful: not negative (96\%), e.g., for tie and men: ``I don't think it's harmful because it's just a fashion choice''; positive stereotype (4\%), e.g., for cake and women: ``cake can be used to describe a woman as sweet and nice looking. For that reason I don't find it harmful.''

\section{Behavior Caption Analysis from Study 3a}
\label{app:caption_results}


Here we present an exploratory analysis of the captions produced by participants for the behavior task in Study 3a.
In Tbl.~\ref{tbl:caption} we compare the captions generated by participants across conditions, and find no statistically significant results.


\begin{table*}
\caption{Descriptive statistics about the captions annotated as a part of Study 3a's behavior measure for the stereotype of women and hair dryer, where toothbrush serves as the control neutral object.}
\label{tbl:caption}
\begin{center}

\begin{tabular}{|>{\centering\arraybackslash}p{2.3cm}|>{\centering\arraybackslash}p{2.2cm}|>{\centering\arraybackslash}p{2.2cm}|>{\centering\arraybackslash}p{1.0cm}|>{\centering\arraybackslash}p{1.0cm}|>{\centering\arraybackslash}p{1.0cm}|>{\centering\arraybackslash}p{1.0cm}|}
    \hline
    \multirow{2}{*}{Condition} &
      \multicolumn{2}{>{\centering\arraybackslash}p{4.4cm}|}{Mention of Hair Dryer / 
      
      Mention of Toothbrush} &
      \multicolumn{2}{>{\centering\arraybackslash}p{2.0cm}|}{Warmth} &
      \multicolumn{2}{>{\centering\arraybackslash}p{2.0cm}|}{Competence} \\ \hline
    Gender of Person Being Described & Women & Men & Women & Men & Women & Men \\
    \hline
    Stereotype-reinforcing &  1.095 (0.679-1.800) & 1.125 (0.679-1.800) & 6.750 $\pm$ 0.386 & 1.107 $\pm$ 0.327 & 1.222 $\pm$ 0.263 & 0.933 $\pm$ 0.225\\
    \hline
    Stereotype-violating & .810 (0.458-1.286) & .857 (0.222-2.500) & 1.182 $\pm$ 0.417 & 0.905 $\pm$ 0.384 & 0.632 $\pm$ 0.265 & 0.750 $\pm$ 0.248\\
    \hline
    Neutral & 0.913 (0.562-1.450) & 0.571 (0.100-1.750) & 1.300 $\pm$ 0.425 & 0.769 $\pm$ 0.300 & 0.783 $\pm$ 0.293 & 1.182 $\pm$ 0.229 \\
    \hline
  \end{tabular}
\end{center}
\end{table*}

\section{Additional Results from Study 3b}
\label{app:3b}
For Study 3b not only did we ask for personal experiential harm from an error, but also societal harm, so that even if the participant does not personally feel harmed, they may feel it on behalf of the stereotyped group. There is a high correlation between the responses to these two versions of the question for each error, which is why we only reported results from personal harm in the main text.

We regress on harm value with the independent variables of stereotype condition, personal harm compared to societal harm, and their interaction effect.
We find that for stereotype-reinforcing vs neutral errors, the coefficient for personal or societal harm is $b = .22$ (95\% $CI$ [-.07, .51], $p=.143$), and that for the interaction effect to be $b = -.18$ (95\% $CI$ [-.59, .23], $p=.398$). For stereotype-reinforcing vs stereotype-violating errors, the coefficient for personal or societal harm is $b = .26$ (95\% $CI$ [-.03, .55], $p=.076$), and that for the interaction effect to be $b=-.22$ (95\% $CI$ [-.63, .19], $p=.288$). In other words, there is no statistically significant difference in the results for societal harm compared to personal harm.


\section{Object Selection from OpenImages}
In OpenImages, of the 600 objects, we select the 20 marked with the most agreement to be stereotypically associated with men, and the 20 marked with the most agreement to be stereotypically associated with women. We then randomly select amongst 20 objects that are marked to have no gender stereotypes associated with them. Left out of this are all human-related categories: \textit{boy}, \textit{girl}, \textit{human eye}, \textit{human face}, \textit{human body}, \textit{human ear}, \textit{human arm}, \textit{human board}, \textit{human hand}; as well as \textit{invertebrate} because there was confusion amongst pilot testers about what this word meant.

The 20 objects stereotyped about men are: \textit{football helmet}, \textit{football}, \textit{cowboy hat}, \textit{hammer}, \textit{sports equipment}, \textit{jet ski}, \textit{truck}, \textit{tie}, \textit{golf ball}, \textit{beer}, \textit{skateboard}, \textit{briefcase}, \textit{plumbing fixture}, \textit{tire}, \textit{wrench}, \textit{suit}, \textit{missile}, \textit{tool}, \textit{rifle}, \textit{shotgun}. The 4 that we consider ``clothing,'' i.e., able to be worn, are \textit{football helmet}, \textit{cowboy hat}, \textit{tie}, \textit{suit}.

The 20 objects stereotyped about women are: \textit{ladybug}, \textit{doll}, \textit{hair spray}, \textit{lily}, \textit{hair dryer}, \textit{perfume}, \textit{kitchenware}, \textit{cat}, \textit{wine glass}, \textit{fashion accessory}, \textit{necklace}, \textit{flower}, \textit{handbag}, \textit{lipstick}, \textit{bathtub}, \textit{face powder}, \textit{cosmetics}, \textit{rose}, \textit{oven}, \textit{brassiere}. The 9 that we consider ``clothing,'' i.e., able to be worn, are \textit{necklace}, \textit{face powder}, \textit{fashion accessory}, \textit{lipstick}, \textit{brassiere}, \textit{cosmetics}, \textit{hair spray}, \textit{handbag}, \textit{perfume}.

The 20 neutral objects are: \textit{pillow}, \textit{owl}, \textit{giraffe}, \textit{balloon}, \textit{jellyfish}, \textit{stop sign}, \textit{french fries}, \textit{eraser}, \textit{shower}, \textit{orange}, \textit{chopsticks}, \textit{window}, \textit{personal flotation device}, \textit{bed}, \textit{goldfish}, \textit{zebra}, \textit{raccoon}, \textit{sea lion}, \textit{microphone}, \textit{popcorn}.

\section{Bias Amplification}
\label{app:biasamp}
Bias amplification is a statistical notion that rests on the idea that any amplification of an existing bias is undesirable, and often used to implicitly capture stereotypes~\cite{zhao2017men, wang2021biasamp, wang2019balanced, hall2022biasamp}.
In this line of work, a ``bias'' is measured in the dataset, e.g., that women are correlated with object A, and so any amplification of this in the model's test-time predictions is considered undesirable, and likely the application of something like a stereotype. This ``bias'' is determined statistically, and two possible formulations come from \citet{zhao2017men} (Bias Amp) and \citet{wang2021biasamp} (Directional Bias Amp).
As an example, \citet{zhao2017men} measures oven, wine glass, and potted plant, to all be biased towards men. From our human annotations, we find all these of these objects to be biased towards women. Thus, mitigation algorithms directed at reducing either of these formulations of bias amplification would actually likely \textit{increase} certain types of harmful errors in an attempt to reduce overall bias amplification.
This formulation also assumes that every label is biased in a way such that one direction of error is worse than another, missing that many labels can be neutral in certain respects, e.g., bowl and table.

We quantify two aspects of each bias amplification metric, which are its abilities to identify either objects as stereotypes (measured by calculating the percentage overlap between the top-$n$ ``biased'' objects and $n$ stereotypes) or the gender direction of the stereotype's alignment (measured by calculating the gender direction on the $n$ stereotyped objects). In Tbl.~\ref{tbl:biasamp} we can see that while both bias amplification metrics are able to approximate the gender that a stereotyped object is correlated with in the COCO dataset reasonably well, this is not true for identifying which objects are stereotypes, nor the gender alignment in the larger OpenImages dataset. Thus, attempts to reduce either metric of bias amplification are likely to inadvertently increase the number of stereotypical errors in an attempt to reduce a ``bias amplification'' error that may not actually be stereotypically harmful.

\begin{table*}
\caption{Comparison of how correlated to human perceptions prior measures of bias amplification, which approximate the directions of bias that are more harmful, are.}
\label{tbl:biasamp}
\begin{center}
\begin{tabular}{ >{\centering\arraybackslash}p{1.8cm} >{\centering\arraybackslash}p{3.5cm} >{\centering\arraybackslash}p{2.cm} >{\centering\arraybackslash}p{2.cm} >{\centering\arraybackslash}p{2.cm} } 
 \hline
Dataset & Metric & Pearson R & Identification of Stereotypes & Alignment of Stereotypes\\
\hline 
COCO & Bias Amp~\cite{zhao2017men} & .5722 & 7/13 (54\%) & 10/13 (77\%) \\
\cmidrule{2-5}
& Directional Bias Amp~\cite{wang2021biasamp} & .6507 & 6/13 (46\%) & 13/13 (100\%) \\
\hline
OpenImages & Bias Amp & .3912 & 124/249 (50\%) & 141/249 (57\%) \\
\cmidrule{2-5}
& Directional Bias Amp & .1502 & 120/249 (48\%) & 153/249 (61\%)\\
\hline
\end{tabular}
\end{center}
\end{table*}

\section{Automatic Discovery of Stereotypes}
Evaluation is sometimes considered secondary to algorithm development, and thus rapid and fully-automated evaluations are often prioritized over those requiring human input. Thus, one might imagine trying to 
automate the determination of which labels are stereotypes, rather than soliciting judgments from human annotators.
To test the limits of this approach, we train a variety of models (Support Vector Machine, Random Forest, and Multi-Layer Perceptron) with hyperparameter search over the number of features and find the highest ROC AUC for predicting whether an OpenImages object is a stereotype given an input of BERT word embeddings~\cite{devlin2019bert} to be 74\%. 
Erroneous predictions include \texttt{carnivore} (is stereotype), \texttt{tennis ball} (is not stereotype), \texttt{infant bed} (is stereotype), \texttt{soap dispenser} (is not stereotype), \texttt{handbag} (is stereotype). 
Given that an object is a stereotype, the highest ROC AUC at predicting which gender is being stereotyped is 85\%. 
Erroneous predictions include \texttt{mixer} (stereotyped with women), \texttt{doughnut} (stereotyped with men), \texttt{houseplant} (stereotyped with women), \texttt{wheel} (stereotyped with men).
These inadequate performance rates indicate that stereotypes are highly contextual, and even with the use of powerful word embeddings which capture bias and social context~\cite{garg2018quantify}, they are insufficient without human input. As we note in the main text, even if the growing power of large language models enables us to predict with higher accuracy which objects are stereotypes, we likely still may want to ensure these annotations come from people themselves~\cite{argyle2023many, hamalainen2023synthetic}, thus allowing room for positionality, explanation, and critical reflection.

\section{Connection to Open-Ended ML Tasks without Correctness}
We scoped our work to machine learning tasks which have a clear notion of error, i.e., ground-truth labels. Here, we consider the implications of our findings for other machine learning tasks which do not have such a clear notion of a error, for example in text generation.
We can also consider the implications of our findings for other machine learning tasks which do not have such a clear notion of a error, for example in text generation. 
Prior work brought to light that word embeddings mirror stereotypes in our society~\cite{bolukbasi2016embeddings, caliskan2017semantics}, such as about occupations and attributes from the Implicit Association Test~\cite{greenwald1998iat}. 
Since then, most follow-up work in this space seeks to remove nearly all gendered associations in text, conflating each such association with harmful ``bias.''
Again we see a similar pattern to the logical fallacy of the harm of one type of error, e.g., a correlation of some set of stereotypical occupations to gender, extending to \textit{all} errors. The nuance is lost when gendered associations of all kinds in word embeddings are equated to stereotypes, and most notions of gender are targeted to be removed from the embeddings. To put this into perspective, in the large body of literature that has followed the discovery of gender biases in the embedding space~\cite{shin2020neutralizing, zhao2018gnglove, karve2019conceptor, dev2019vector, kaneko2019debiasing, ravfogel2020null, dev2020embeddings, manzini2019multiclass}, all eight of these works would, as far as we can tell, attempt just as much to debias words like ``table'' and ``apple'' as they would ``homemaker'' and ``doll.'' While it is not clear what exactly is the desired state of debiasing (e.g., describing the world as it is, prescribing the world as it ought to be, aligning with people's existing stereotypic expectations~\cite{wang2021acceptance}) it surely seems that more thinking should be done on the different implications of debiasing stereotypes as opposed to debiasing more neutral concepts.

\end{document}